 \documentclass [aps,pra, onecolumn,a4paper]{revtex4}

 \pdfoutput=1

\usepackage{amssymb,amsmath,amsfonts,graphicx}
\usepackage{bm}
\usepackage{epsfig,color,times}
 \usepackage{bm,color}
\usepackage{graphicx}
\usepackage{amssymb}
\usepackage{epstopdf}

\begin{document}

\title{Turbulence in a wedge: the case of the mixing layer}
\author{Yves Pomeau and Martine Le Berre}
\affiliation{ 
Laboratoire d'Hydrodynamique, Ladhyx, (CNRS UMR 7646), Ecole Polytechnique, 91128 Palaiseau, France }

%\author{Yves Pomeau$^1$ and Martine Le Berre$^2$}
%\affiliation{ $^1$ Ladhyx (CNRS UMR 7646), Ecole Polytechnique, 91128 Palaiseau, France 
%\\$^2$  Ismo (CNRS UMR 8214), Universit\'{e} de Paris-Saclay,
%91405 Orsay, Fra

\date{\today}

\begin{abstract}
The ultimate goal of a sound theory of turbulence in fluids is to close in a rational way the Reynolds equations, namely to express the tensor of turbulent stress as a function of the time average of the velocity field. Based on the idea  that dissipation in  fully developed turbulence is by singular events resulting from an evolution described by the Euler equations, it has been recently observed that the closure problem is strongly restricted, and that it implies that the turbulent stress is a non local function in space of the average velocity field, a kind of extension of classical Boussinesq theory of turbulent viscosity.  This leads to rather complex  nonlinear integral equation(s) for the time averaged velocity field. This one satisfies some symmetries of the Euler equations. 
Such symmetries were used by Prandtl and Landau to make various predictions about the shape of the turbulent domain in simple geometries. We explore specifically the case of mixing layer  for which  the average velocity field  only depends on the angle in the wedge behind the  splitter plate.  This solution yields a pressure difference between the two sides of the splitter which contributes to the lift felt by the plate. Moreover, because of the structure of the equations for the turbulent stress, one can satisfy the Cauchy-Schwarz inequalities, also called the realizability conditions, for this turbulent stress. Such realizability conditions cannot be satisfied with a simple turbulent viscosity.   

\end{abstract} 

\maketitle

\section{Introduction}
\label{Introduction}
One fundamental result in fluid mechanics goes back to Newton's Principia and states that at large constant velocity (large Reynolds number in modern terms) the drag  force on a blunt body is proportional to the square of its velocity times its cross section times the mass density of the fluid. This remarkable result is not trivial because it is fully independent on the viscosity that is-a priori- responsible of dissipation in fluids. An explanation is that dissipation takes place in singular events  \cite{leray} resulting from the evolution described by Euler inviscid equations. Although viscosity becomes relevant in the final stage of this evolution the amount of energy dissipated there, is  independent of the viscosity because  it is the energy present both initially and in this final stage of the singular solution which is fully described by the energy conserving Euler dynamics \cite{YM}-\cite{NLS}.  This explanation is the one we adopt here following  reference \cite{chaos} where it was  shown that this  approach leads to an expression of turbulent stress tensor formulated in terms of the time-averaged velocity field, which is non local in space. The non locality  follows  from the constraint that there is no  physical parameter, like a length or a velocity, independent of the average velocity field. This constraint is the key leading to our model for the Reynolds stress  tensor  defined by  the correlation of the velocity fluctuations ${\bold{U'}}$ by  the relation (neglecting the contribution of the viscous stress)  
  \begin{equation} 
\sigma_{ij}^{Re}({\bold{X}}) =  \rho <  U'_i ({\bold{X}}) U'_j ({\bold{X}}) >
 \label{eq:sigRe1}
\end{equation}  
where $<\cdot>$  mean a time  average.  Defining  $U({\bold{X}})$   (without bracket in order to lighten the writing) as the  time average of the velocity,
 our  non local model belongs to the class of equations written as the sum of two contributions
 \begin{equation} 
\sigma_{ij}^{Re}({\bold{X}}) = \tilde \sigma_{ij}^{Re}({\bold{X}})  \, + \, \sigma_{ij}^{Re,p}({\bold{X}}) 
 \label{eq:sig12}
\end{equation}  
where the first term is the non diagonal tensor
 \begin{equation} 
 \tilde \sigma_{ij}^{Re}({\bold{X}}) =\tilde \gamma \rho \vert  {\bold{\nabla}}\times {\bold{U}}({\bold{X}})\vert^{1-\alpha} \int {\mathrm{d}} {\bold{X}}' \vert{\bold{\nabla}}\times {\bold{U}}({\bold{X'}})\vert^{\alpha} \left (\frac{1}{\vert{\bold{X}} - {\bold{X}}' \vert }- \frac{1}{\vert {\bold{X}}' \vert} \right)   (U_{i, j} + U_{j,i})({\bold{X}}') 
 \label{eq:sig11}
\end{equation}  
and the second term in the r.h.s. of (\ref{eq:sig12}) is the diagonal tensor
 \begin{equation} 
\sigma_{ij}^{Re, p}({\bold{X}}) =  \delta_{ij}  \gamma \rho  \vert {\bold{\nabla}}\times {\bold{U}}({\bold{X}}) \vert ^{1-\alpha }\int {\mathrm{d}} {\bold{X}}'   \vert {\bold{\nabla}}\times {\bold{U}}({\bold{X'}}) \vert  ^{\alpha + 1} \left (\frac{1}{\vert{\bold{X}} - {\bold{X}}' \vert }- \frac{1}{\vert {\bold{X}}' \vert} \right).
 \label{eq:sig22}
\end{equation}
In (\ref{eq:sig11}) and (\ref{eq:sig22})  the exponent $\alpha$ is such that $\vert \alpha \vert <1$, $\tilde \gamma$  and $\gamma$ are dimensionless constants. Those three quantities
 have to be found either by analyzing experimental results and/or numerical simulations.  In  equation  (\ref{eq:sig12}) the indices $i$ and $j$ are for the Cartesian coordinates. They should not be confused with the two indices $1$ and $2$ attributed to the two sides of the mixing layer later in this paper. Above and elsewhere we shall use the notation with comma  in the subscript to denote derivation, so that $U_{i, j} $ is for $\frac{\partial U_i}{\partial X_j}$. 

Let us explain the way the  Reynolds stress  $\sigma_{ij}^{Re}$  is built.  As one can check it has the same scaling properties as the turbulent stress imagined long ago by Boussinesq  \cite{schmitt}, namely it scales like velocity square times  $\rho$ times a length (called now the Prandtl length scale). However it has some features requiring to be explained. The first obvious feature of this equation is that it is obviously not invariant under spatial translation, because the counter term $ \frac{1}{\vert {\bold{X}}''\vert}$ in the integral kernel introduces an (unspecified) origin of coordinates from which the vector  ${\bold{X}}' $ is measured. This breaking of the translational invariance is not surprising by itself because the turbulent Reynolds stress depends on the average properties of the turbulent fluctuations (see below) which depend on the geometry of the walls limiting the fluid. For general shape of the walls bounding the fluid an extended version of the integral kernel is to take the Green's function of the Laplace operator with Dirichlet boundary conditions. This would amount to replace in equations  (\ref{eq:sig11})   and  (\ref{eq:sig22})  the integral kernel $\left (\frac{1}{\vert{\bold{X}} - {\bold{X}}' \vert }- \frac{1}{\vert {\bold{X}}' \vert} \right)$ by $K({\bold{X}}, {\bold{X}}')$ where $K$ is solution of Laplace's equation with respect to the variable  ${\bold{X}}'$  
\begin{equation} 
 \nabla^2 K({\bold{X}}, {\bold{X}}') =  \delta_{\textsc{D}} ({\bold{X}} - {\bold{X}}')
  \label{eq:sigLap}
\end{equation} 
where $  \delta_{\textsc{D}}$ is Dirac's delta function. We shall deal with the mixing layer  set up where a half infinite splitting plate ends up on a line of Cartesian equation $x = y = 0$, see Fig.\ref{fig:schema}. Because this mixing layer has a simple geometry it is natural to take as origin of coordinates a point on the edge of the splitter, as we shall do. This is justified by the fact that, without this counter term in the integral equation, the integral diverges logarithmically when done with respect to the variable $z$, along the edge of the splitter. Subtracting this counter term one finds a converging result because the divergence of the two terms  cancel each other. Moreover, including this counter term, the turbulent stress scales like the product of $\rho$ by the square of a velocity square. More complex physical situations like the one of a turbulent flow around an obstacle like a sphere or a turbulent Poiseuille pipe flow  require to introduce the more complex integral kernel $ K({\bold{X}}, {\bold{X}}')$ just introduced. 

 Another property of the expressions (\ref{eq:sig11}) and   (\ref{eq:sig22}) for the turbulent stress is the explicit occurrence of the vorticity. Vorticity is known to play a central role in non homogenous and non isotropic turbulence because once vorticity is present it is amplified by vortex stretching. Moreover  our model agrees with Landau's description  of wakes (section 35 in  \cite{LL}), made up of two domains, one potential and the other  rotational. In the rotational -non potential- domain there is a kind of equilibrium on average between the growth of vorticity, by vortex stretching and by injection from the boundaries, and its damping in singular events.  About the potential domain, Landau  states that '' outside the region of rotational flow the turbulent eddies must be damped and must be so more rapidly  for small eddies which do not penetrate far away in the  potential domain".

The expression (\ref{eq:sig12}) agrees with the basic constraints derived from the structure of Euler fluid equations, except the one of  reversibility (which constrains smooth solutions but not singular ones). Irreversibility in  the first term (\ref{eq:sig11})  is reflected by the fact that the product of absolute value of the vorticity   by the strain tensor  components makes the turbulent stress $ \tilde \sigma_{ij}^{Re}$ change sign when changing the sign of $U$, whereas the inertial stress $\rho U_i U_j$    and $\sigma_{ij}^{Re, p}$ do not change sign. In our picture of turbulence the irreversibility is due to the evolution toward finite time singularities of the Leray type \cite{leray}, the solution  disappearing close to the collapse time due to molecular dissipation at small scales, so that  dissipation will ultimately yield the friction of Newton's drag law.    This picture of dissipation  at collapse time is analogous to Maxwell's theory  of molecular viscosity of gases  \cite{Mxw}, where the velocity difference between two colliding particles induced by a macroscopic shear flow reduces to zero when the particles collide, transforming the energy of this velocity difference into heat.   As just written,  $\sigma_{ij}^{Re, p}$ does not change sign as ${\bold{U}}$ changes sign. 
Therefore the addition of $\sigma^{Re, p}$ to turbulent stress will represent a contribution to this stress that does not participate to friction, that is possible a priori. In particular in the geometry of the mixing layer the $\sigma_{zz}^{Re}$ component of the stress cannot enter into the friction which, by symmetry, is a force directed along the $x$ axis and so depend on the components of $\sigma_{ij}^{Re}$ with at least one index $i$ or $j$  equal to $x$, which excludes $\sigma_{zz}^{Re}$. 

 Let us now explain  why we add the diagonal tensor $\sigma_{ij}^{Re, p}$ to the tensor $\tilde \sigma_{ij}^{Re}$.                                                                                                                                                                                                                                                                                                                                                                                                                                                                                                                                                                                                                                                                                                                                                                                                                                                                                                                                                                                                                                                                                                                                                                                                                                                                                                                                                                                                                                                                                                                                                                                                                                                                                                                                                                                                                                                                                                                                                                                                                                                                                                                                                                                                                                                                                                                                                                                                                                                                                                                                                                                                                                                                                                                                                                                                                                                                                                                                                                                                                                                                                                                                                                                                                                                                                                                                                                                                                                                                                                                                                                                                                                                                                                                                                                                                                                                                                                                                                                                                                                                                                                                                                                                                                                                                                                                                                                                                                                                                                                                                                                                                              
In many theories of turbulence, going back to Boussinesq and to Reynolds \cite{schmitt},  the turbulent stress tensor is taken as proportional to the rate of strain, namely to the tensor  $\tau_{ij}= (U_{i, j} + U_{j,i})$. Therefore the trace of  $\tilde \sigma_{ij}^{Re}$  is proportional to the divergence of the velocity field,  ${\bold{\nabla}}\cdot {\bold{U}} = U_{i, i}$ (with summation on repeated indices), which is zero for incompressible fluids. The  null trace  of   the tensor  $\tilde \sigma_{ij}^{Re}$ is not compatible with the definition of  the Reynolds stress in (\ref{eq:sigRe1}) which implies that all diagonal elements must be positive,  and must be related to the off-diagonal elements by the Schwarz inequality,  these two conditions being named realizability conditions in the turbulence community \cite{sch}, see (\ref{eq:appDcondreal}) in Appendix \ref{sec:app-sigI}. 

In our model the   diagonal tensor    $\sigma_{ij}^{Re, p}$  plays the role of a time averaged pressure (but spatially depending) caused by the vorticity, that  we call {\textit{turbulent pressure}}  which has to be added to the usual time averaged pressure (also spatially depending)  which exists without vorticity. Recall that  in  dynamical compressible systems the couple $({\bold{U}},p)$ is not unique, because $p$ is  a scalar  jauge field defined up to an additive constant.  In other words the  pressure is not  an independent variable, but a Lagrangian multiplier necessary to ensure the compressibility, since it fulfills the relation ${\bold{\nabla}} p = \Delta {\bold{U}}$.
 We show below that the realizability conditions  
are fulfilled by the addition of the diagonal tensor  $\sigma_{ij}^{Re, p}$. In the  mixing layer  case  with quasi-equal  incoming velocities  the conditions are fulfilled by taking equal coefficients $\tilde \gamma = \gamma$  in (\ref{eq:sig11})  and (\ref{eq:sig22}). In other cases  we expect that the factor $\gamma$ in turbulent pressure can be adjusted to satisfy the constraints of the realizability.

In summary, our model of the Reynolds stress contains a part  which is linear with respect to the strain tensor $\tau_{ij}$ plus the turbulent pressure  reflecting the role of   the vorticity,  
\begin{equation}
p^{turb} = \sigma_{ii}^{Re,p}.
\label{eq:pturb}
\end{equation}
The source of this turbulent pressure  is the square of the vorticity. That this pressure is linked to vorticity is also in agreement with the fact that turbulence is characterized by vorticity in real turbulent flows. 
 Note that it is well-known that vorticity is a source of low pressure in incompressible fluids, irrespective of the sign of the vorticity, which makes the turbulent wake domain to  suck part of the flow of the potential domain, leading for exemple to the Coanda effect \cite{coanda}. The interest of introducing this turbulent pressure will hopefully be more obvious in the case of the turbulent mixing layer studied below. There we solve the equation for the balance of momentum by eliminating the scalar pressure, that allows to obtain an analytical expression for the  average velocity field. In a second step  we set an expression for the total scalar pressure 
 \begin{equation}
 p_{t}= p^{turb} +p 
 \label{eq:ptotal}
\end{equation}
  including the average of the standard pressure $p$ (associated to  the so-called RANS equation) and the turbulent pressure.  The latter is computed 
 by using (\ref{eq:sig22}), and  the standard pressure is the difference between the total pressure and the turbulent pressure.

Below we  consider the  case of the mixing layer where the equations written in polar coordinates depend on one variable only, the angle. 
Section \ref{sec:Formulae} is devoted to the relatively non trivial task of writing equation (\ref{eq:sig11}) fully explicitly, and to derive the equation for the balance of the total stress $\Sigma_{ij}$ resulting from it, this tensor being the sum of the turbulent stress,  the pressure and  the inertial stress, see (\ref{eq:sigg}). In sec.\ref{sec:Small}  we solve this problem for a small velocity difference of the two merging flows.

\section{Formulae in polar coordinate}
\label{sec:Formulae}
In this section we derive in polar coordinates the explicit equation for the balance of stress. The whole calculation is fairly complex and is done by using mixed coordinates, polar coordinates for the argument of the functions and Cartesian coordinates for the velocity field and the stress tensor. 

\subsection{Stress balance in polar coordinate}
\label{sec:Formulae.1}

\begin{figure}
\centerline{ 
\includegraphics[height=1.5in]{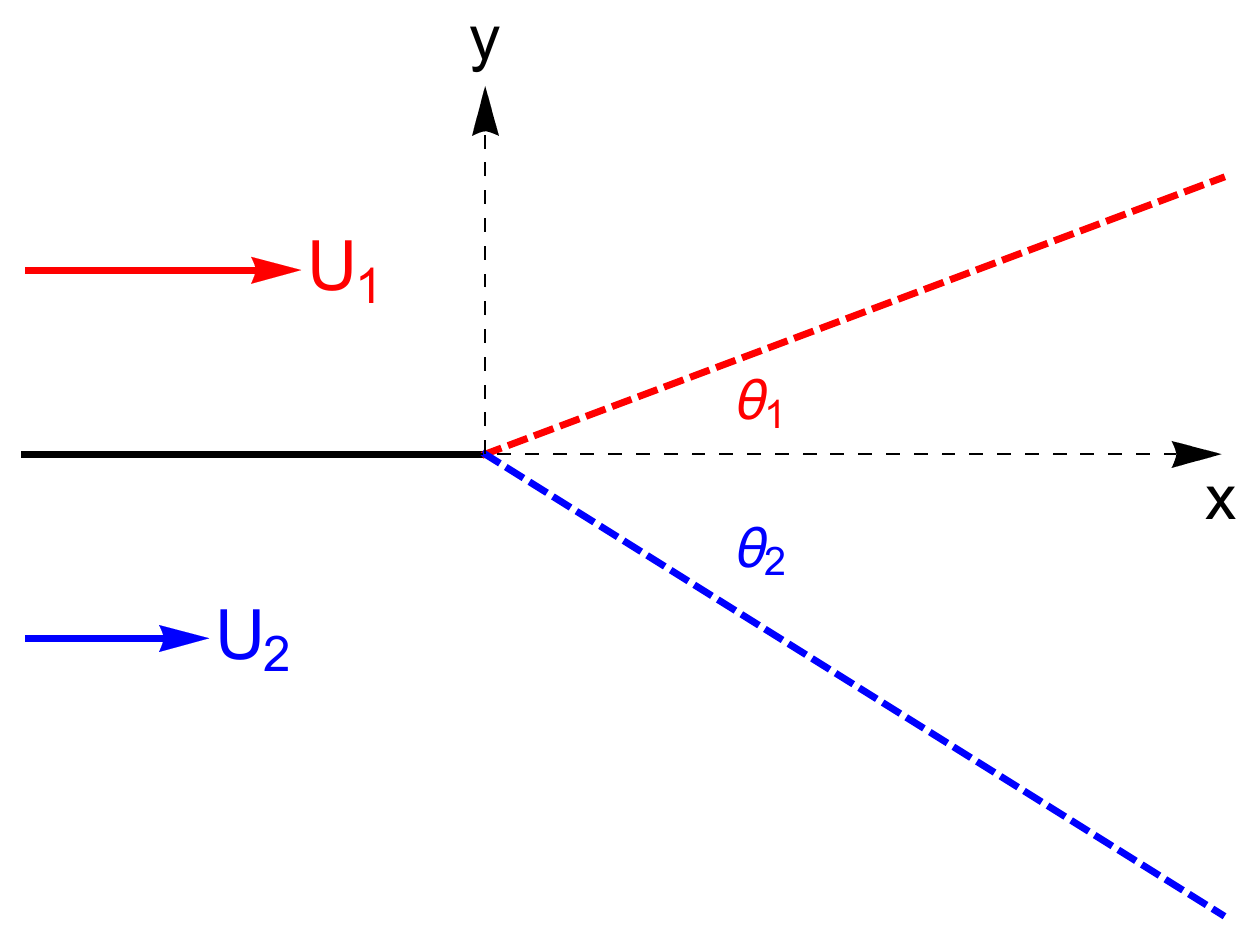}
  }
\caption{ Cross section a mixing layer in the plane $(x,y)$. The  obstacle is a semi-infinite plane, located in the domain ($x<0,  -\infty <z < \infty$),  the  incident flow has different velocities  amplitude $U_{1}$ and $U_{2}$ above and below the obstacle.  Behind the obstacle,  a turbulent wedge is formed,   materialized by  the two rays making angles $\theta_{1}$ and $ \theta_{2}$ with the $x$-axis (in Landau's description   the flow is non potential inside this wedge).}
\label{fig:schema}
\end{figure}

We consider the turbulent wake behind a semi-infinite plane  board (the splitter) supposed to be  horizontal  in the plane $(x,z)$, limited to the domain $x<0$, and submitted to two  inviscid parallel flows with different velocities in the $x$ direction, the upper one with velocity ${\bold{U}}_{1}$ and the lower one with velocity ${\bold{U}}_{2}$, schematized in fig.\ref{fig:schema}. 

Because of symmetries of the equations and of the geometry the time-average of the velocity depends on the angle only, which leads to use cylindrical coordinates $(r, \theta, z)$.  The present derivation can be extended to other flow geometries where the velocity field depends on an angle only  like, for instance,  a uniform parallel flow impinging an inclined plate at high Reynolds number.  This assumption of a large Reynolds number implies that we consider what happens at distances from the splitter large enough to make large the corresponding Reynolds number.

In the situations to be considered no fluid parameter depends on the coordinate $z$ perpendicular to the plane $(x, y)$. Moreover, in the limit we consider, viscous effects as negligible so that, as had been shown by Prandtl, the average velocity in this plane depends on the polar angle $ \theta$ only, with $\theta$ increasing from  $ \theta = 0 $ on the $x$ axis, to $ \pi/2$ on the vertical axis $y$, to $\pi$ on the upper part of the plate and  symmetrically to  $-\pi$ on the lower part. 
Setting $ \theta = 0 $ on the $x$ axis, the coordinates $x$ and $y$ are related to the angle $\theta$   and the radius $r$ by
 \begin{equation} 
  x = r \cos \theta
  \label{eq:x}
\end{equation}
  \begin{equation} 
   y = r \sin \theta
   \label{eq:y}
\end{equation}
  
The incompressible time-averaged  velocity field ${\bold{U}}$ is in the plane $(x,y)$ and is given by the stream function $ \Psi = r g(\theta)$, where the function $g(\theta)$ is to be found. Let the Cartesian components of the velocity  ${\bold{U}}$  be  $U_{x}= u$ and $U_{y}= v$ in the direction $x$ and $y$ respectively. From $u =  - \Psi_{,y}$ and $v = \Psi_{,x} $ where comma are for partial derivatives, one has
 \begin{equation} 
 u =  - (g \sin \theta  + g' \cos \theta)
  \label{eq:u}
\end{equation}
  \begin{equation} 
  v =  g \cos \theta  -  g' \sin \theta 
    \label{eq:v}
\end{equation}
where $g' =   \frac{dg}{d\theta}$. Hopefully, no confusion will  arise between  this symbol of derivation and  the primed notation $X'$ for coordinates  in  (\ref{eq:sig11})-(\ref{eq:sig22}) and below in (\ref{eq:sigint}).
The $z$ component of the curl of the velocity field is  the only non vanishing component of the vorticity given by  
 \begin{equation} 
  {\bold{\nabla}}\times {\bold{U}}({\bold{X}})=  \frac{1}{r} (g + g'') {\bold{e}}_{z},
      \label{eq:curl}
\end{equation}
 where $g'' = \frac{d^2g}{d\theta^2}$ and ${\bold{e}}_{z}$ is the unit vector along $z$. 

The integration on the coordinate $z'$ in equations (\ref{eq:sig11})-(\ref{eq:sig22}) can be performed because the variable $z'$ occurs only in the denominators in $\frac{1}{\vert  {\bold{X}}-  {\bold{X}}' \vert }- \frac{1}{\vert  {\bold{X}}' \vert}$.
The result is
 \begin{equation} 
 \int {\mathrm{d}} z'  \left(\frac{1}{\vert {\bold{X}} -  {\bold{X}}' \vert }- \frac{1}{\vert  {\bold{X}}' \vert} \right) = \ln \left[\frac{a(r,r',\theta - \theta')}{a(0,r',\theta - \theta')}\right] ^{2}
   \label{eq:sigint}
\end{equation}   
 
where 
  $$ a^2(r,r',\theta - \theta') = r^{2}+r'^{2}-2rr'\cos(\theta - \theta')$$
  
  The next step in this calculation is to write explicitly the condition of balance of momentum. Let us define the full stress tensor $ \Sigma_{ij}$,  which  is the sum of three terms involving the contribution of inertia, of the Reynolds stress and of the pressure $p   \delta_{ij}$ with $\delta_{ij}$ Kronecker symbol,
   \begin{equation} 
   \Sigma_{ij} = \rho \, U_i U_j +  \sigma_{ij}^{Re} + p   \delta_{ij}. 
    \label{eq:sigg}
\end{equation}   
Within our model (\ref{eq:sig12})  it can also be written in the form
  \begin{equation} 
   \Sigma_{ij} = \rho \, U_i U_j + \tilde \sigma_{ij}^{Re} + (p+p^{turb} )  \delta_{ij} 
    \label{eq:sigg2}
\end{equation}   
which is the one  used  below in order to  get the expression of the  average velocity as a function of $\theta$.
In  the mixing layer case the balance of momentum  does not depend on $z$. In cartesian coordinates the balance  is given by  the two conditions
     \begin{equation} 
  \Sigma_{xx, x} +  \Sigma_{xy, y}  = 0
    \label{eq:sig1}
\end{equation}   
and
   \begin{equation} 
 \Sigma_{yy, y} +  \Sigma_{xy, x}  = 0.
    \label{eq:sig2}
\end{equation}   
which reduce,  in polar coordinates, to  the two following ODE's (ordinary difference equation) with respect to the variable $\theta$
     \begin{equation} 
      - (\sin \theta ) \Sigma_{xx, \theta} +  (\cos \theta) \Sigma_{xy, \theta}  = 0
         \label{eq:sig10}
\end{equation}   
    \begin{equation} 
 (\cos \theta ) \Sigma_{yy, \theta} - (\sin \theta) \Sigma_{xy, \theta}  = 0
         \label{eq:sig20}
\end{equation}   
Up to a global multiplication by the mass density $\rho$  that will not be written explicitly, 
let us consider the contributions of the first two terms in (\ref{eq:sigg2}).
  The contribution of $U_i U_j$ is expressed simply in terms of the stream function as 
  \begin{equation} 
U^2_x = u^2  = (g \sin \theta +  g'\cos \theta)^2
         \label{eq:u2x}
\end{equation}   
  \begin{equation} 
U^2_y = v^2 = ( g\cos \theta -  g'\sin \theta)^2
         \label{eq:u2y}
\end{equation}  
  \begin{equation} 
U_x U_y = u v= \sin \theta \cos \theta (g'^2 - g^2) + g g' ( \sin^2 \theta -  \cos^2\theta) 
         \label{eq:uxy}
\end{equation}   
  At this step the unknown functions are $g(.)$ and the pressure $p$, depending both on $\theta$ only. It is possible to eliminate the pressure by taking the curl of the two ODE's for the stress tensor, as usually done in this kind of problem. However in the present case one more step can be made because the tensor $ \Sigma_{ij}$ and the pressure $p$ depend on $\theta$ only. Because of that the pressure appears by its derivative with respect to $\theta$ only in equations (\ref{eq:sig10}) and  (\ref{eq:sig20}). Therefore one can eliminate the pressure from those equations by algebraic handling only, without increasing the order of derivation in the final equation. Let us define $ \tilde{\Sigma}_{ij}$ as $ \Sigma_{ij}$ without the total pressure term defined in (\ref{eq:ptotal})
   \begin{equation} 
 \Sigma_{ij} = \tilde{\Sigma}_{ij} + p_{t} \delta_{ij}.
  \label{eq:Sigtild}
\end{equation}
 After straightforward algebra the components of $ \tilde{\Sigma}_{ij}$ satisfy the following single equation without the pressure and with $g(\theta)$ only as unknown function 
    \begin{equation} 
     \sin \theta  \cos \theta ( \tilde{\Sigma}_{yy, \theta} -  \tilde{\Sigma}_{xx, \theta}) + ( \cos^2 \theta -  \sin^2 \theta )  \tilde{\Sigma}_{xy, \theta} = 0
    \label{eq:sig3}
\end{equation} 
which is the basic equation to be solved. Recall that
the stress  $\tilde{\Sigma}_{ij}$ is the sum of  the inertial term $U_i U_j$ plus the stress  tensor $\tilde \sigma_{ij}^{Re}$ given in equation  (\ref{eq:sig11}) both being  a function of the time-averaged velocity field and ultimately of the stream function of the same averaged velocity $\Psi = r g(\theta)$, $g(.)$ being the unknown function to be found by solving equation (\ref{eq:sig3}). From the computational point of view, the above equation (\ref{eq:sig3}) is formally independent of the isotropic part of the stress, coming from the pressure.  
Either equation   (\ref{eq:sig10}) or  (\ref{eq:sig20}) can be used to find the total pressure, both equations being compatible because of the way equation  (\ref{eq:sig3}) is derived.  Using polar coordinates (\ref{eq:sig10}) and  (\ref{eq:sig20})  lead to the relation
   \begin{equation} 
  \frac{1}{\rho} ( p_{t})_{,\theta} = - \sin^2 \theta \; \tilde{\Sigma}_{xx, \theta} -  \cos^2 \theta\;  \tilde{\Sigma}_{yy, \theta} + 2 \sin \theta  \cos \theta \; \tilde{\Sigma}_{xy, \theta} 
    \label{eq:ptheta}
\end{equation}

Looking at the literature on the theory of the mixing layer one finds often a somewhat expeditious treatment of the pressure gradient which is set rather arbitrarily to zero. This seems not justified at least for a number of reasons. First pressure in the equations of incompressible fluid mechanics is necessary to impose incompressibility. In the present problem, if $p_{,\theta}$ is set to zero arbitrarily, there is a conflict because one has the two equations (\ref{eq:sig10}) and  (\ref{eq:sig20}) for one unknown function ($g(.)$ here). Neglecting the pressure  term is also unphysical because this pressure depends on $\theta$  in such a way that it tends to different values as $\theta$ tends to $\pi$ and $-\pi$. This non zero pressure difference yields the lift force on the semi infinite plate that is the integral along the surface of the plate of the pressure difference given by the following expression 
 \begin{equation} 
 p_{t}(\pi ) -  p_{t}(- \pi ) =  \int_{- \pi}^{\pi} {\mathrm{d}} \theta \;  (p_{t})_{,\theta}
     \label{eq:sig3p}
\end{equation} 
where $ (p_{t})_{,\theta}$ is given by (\ref{eq:ptheta}).
Furthermore this pressure difference is also needed to balance the loss of energy in the turbulent mixing layer.

\subsection{ Stress tensor $\tilde \sigma_{ij}$}
\label{sec:Formulae I}

As defined in (\ref{eq:sig11})  the stress tensor $\tilde \sigma_{ij}$ depends (linearly) on the rate of strain tensor  $\tau_{ij}$ defined as
 \begin{equation} 
 \tau_{ij} = U_{i, j} + U_{j,i}.
  \label{eq:tauij}
\end{equation}   
The components of the  strain tensor  $\tau_{ij}$ are expressed in function of the stream function as
 \begin{equation} 
\tau_{xx} = \frac{2}{r'}  \sin \theta'  \cos \theta' ( g + g''), =-\tau_{yy}
  \label{eq:tauxx}
\end{equation}
  \begin{equation} 
\tau_{xy} = \frac{ \sin^2 \theta' - \cos^2 \theta'}{r'} ( g + g'')
    \label{eq:tauxy}
\end{equation}
That $\tau_{yy} + \tau_{xx} = 0 $ is a straight consequence of the incompressibility in 2D.  As explained in the introduction, it shows that the contribution $\tilde \sigma_{ij}$ to the stress tensor  $\sigma_{ij}$ does not meet the requirement of realizability by itself, because the trace of the Reynolds stress has to be positive, whereas the tensor $\tilde \sigma_{ij}$ has a vanishing trace.  Recall that the diagonal tensor $\sigma_{ij}^{Re,p}$ called turbulent pressure has been added  to $\tilde \sigma_{ij}^{Re}$ in (\ref{eq:sig12})  in order to correct this point.
%As we argued before this constraint of realizability disappears if dissipation and momentum transfer in highly turbulent flows is by singular solutions of the Euler equations, as we assume.  
 
To lighten the coming algebra, let us introduce a new tensor $\tilde\tau_{ij}$ defined by
 \begin{equation} 
\tau_{ij} =  \frac{1}{r}\tilde\tau_{ij} (\theta)
  \label{eq:tautild}
\end{equation}
in order to split the integrals in equation (\ref{eq:sig11}), and (\ref{eq:sig22}), into a part involving the angle only, times the result of an integral on $r'$ that can be carried explicitly. The problem of writing explicitly the momentum balance  is now reduced to an integral equation for $\theta$-dependent functions only. 

\subsubsection{momentum balance}
\label{sec: balance I}
Because the dependence on $z$ and $z'$ is only in the denominators in $ \left (\frac{1}{\vert{\bold{X}} - {\bold{X}}' \vert }- \frac{1}{\vert {\bold{X}}' \vert} \right)$ the integral over $z'$ can be carried explicitly. 
  Now concerning   the integration over the variable $r'$  in  the definition  of  $ \tilde \sigma_{ij}^{Re}( {\bold{X}})$ , 
the integral  written in  equation (\ref{eq:sig11})  reduces  to $   \vert (g + g'')(\theta') \vert^{\alpha} \tilde\tau_{ij}(\theta' ) \int _0^{\infty} \mathrm{d} r' r'^{-\alpha} \ln(1 +  (r/r')^{2}- 2 (r/r') \cos(\theta - \theta'))$  where $\tilde \tau$ is defined in (\ref{eq:tautild}). Setting $\zeta=r'/r$ allows to get rid of the variable $r$, we get 
 \begin{equation} 
 \tilde  \sigma_{ij}^{Re}( {\bold{X}}) =   \rho \, \tilde \gamma \vert (g + g'') \vert^{1-\alpha} \int {\mathrm{d}} \theta'\,  \tilde\tau_{ij}(\theta')  \, \vert (g + g'')(\theta') \vert^{\alpha} {\mathcal I}(\theta -  \theta'), 
   \label{eq:sigfin}
\end{equation}   
where the kernel ${\mathcal I}$ is defined by the relation
 \begin{equation} 
 {\mathcal I} (\theta-\theta') =  \int _0^{\infty} \frac{{\mathrm{d}} \zeta}{\zeta^{\alpha}} \ln(\frac{1 +  \zeta^2 - 2 \zeta \cos(\theta - \theta')}{\zeta^2}), 
    \label{eq:defI}
\end{equation}   
This integral converges if $0 < \alpha < 1$ as assumed. Integrating by parts one obtains
 \begin{equation} 
 {\mathcal I}(\theta-\theta') = \frac{2}{1-\alpha }  \int _0^{\infty} \frac{\zeta^{-\alpha} {\mathrm{d}} \zeta }{1 +  \zeta^2 - 2 \zeta \cos(\theta - \theta')} (1- \zeta\cos(\theta - \theta') )
     \label{eq:defI2}
\end{equation}   

The equation to be satisfied by $g(.)$  will be deduced by putting all the above results into equation (\ref{eq:sig3}). Consider first the contribution of the term of inertia, namely the tensor $U_i U_j$ when dropping the factor $\rho$. Its contribution $ C$ to the left hand side of equation (\ref{eq:sig3}) can be written as a quantity quadratic with respect to $g$ and its derivatives.  All calculations done yield the simple looking result, 
  \begin{equation}  
C = -g(g+g''),
   \label{eq:sigfinal1}
\end{equation} 
see Appendix A for details.
Let call $ D$ the other contribution  to the left-hand side of equation (\ref{eq:sig3}) which comes from the turbulent stress. Using the expressions given above, one obtains
   \begin{equation}  
 D = - \tilde \gamma  \frac{{\mathrm{d}}}{{\mathrm{d}}\theta} \left(\vert (g + g'')(\theta) \vert^{1-\alpha} \int_{-\pi}^{\pi}  {\mathrm{d}} \theta' {\mathcal I}(\theta -  \theta') \right) \vert (g + g'')(\theta') \vert^\alpha (g + g'')(\theta') \cos 2(\theta-\theta') 
   \label{eq:sigfinal 2}
\end{equation} 
as detailed in Appendix B.
The angular integral is carried over the full angle, namely from $-\pi$ to $\pi$, even though the mixing layer is expected to be concentrated near $\theta = 0$. However we assume that the perturbation to the incoming potential flow which extends itself to the full angular domain, has a very small amplitude far from the angular wedge of the mixing layer. This view is confirmed by the calculation done in section \ref{sec:Small} in the limit of a small velocity difference between the two sides of the mixing layer. Looking at the literature it is not obvious to see if a strictly bounded turbulent wedge is predicted, which poses the problem of the condition across the limit of this region or if a smooth continuity exists between the potential flow and the turbulent layer. Practically the matter is not that meaningful  since  a turbulent domain penetrating into the potential flow with an exponentially decaying amplitude does not make much difference with an exactly bounded non potential domain.

\begin{figure}
\centerline{ 
%(a)
(a)\includegraphics[height=1.5in]{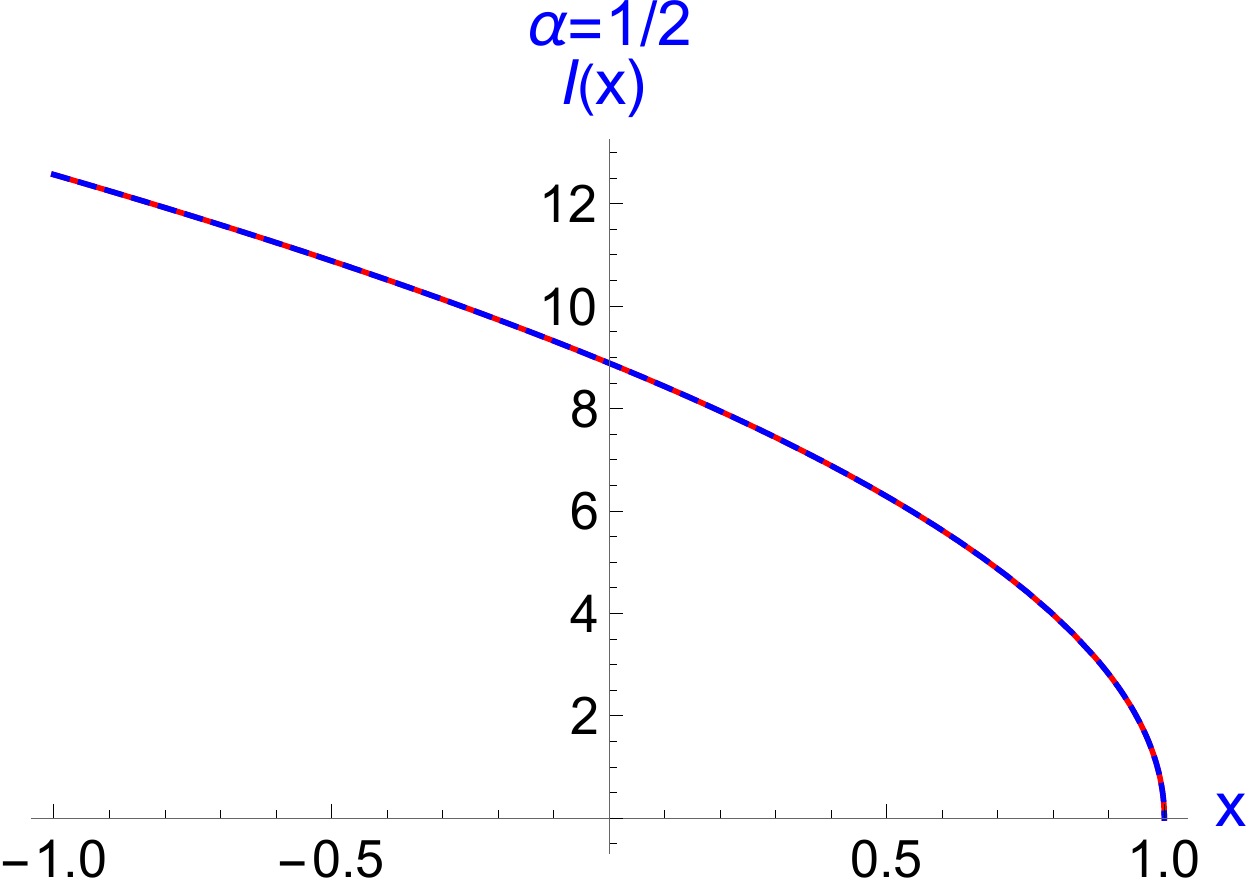}
(b)\includegraphics[height=1.5in]{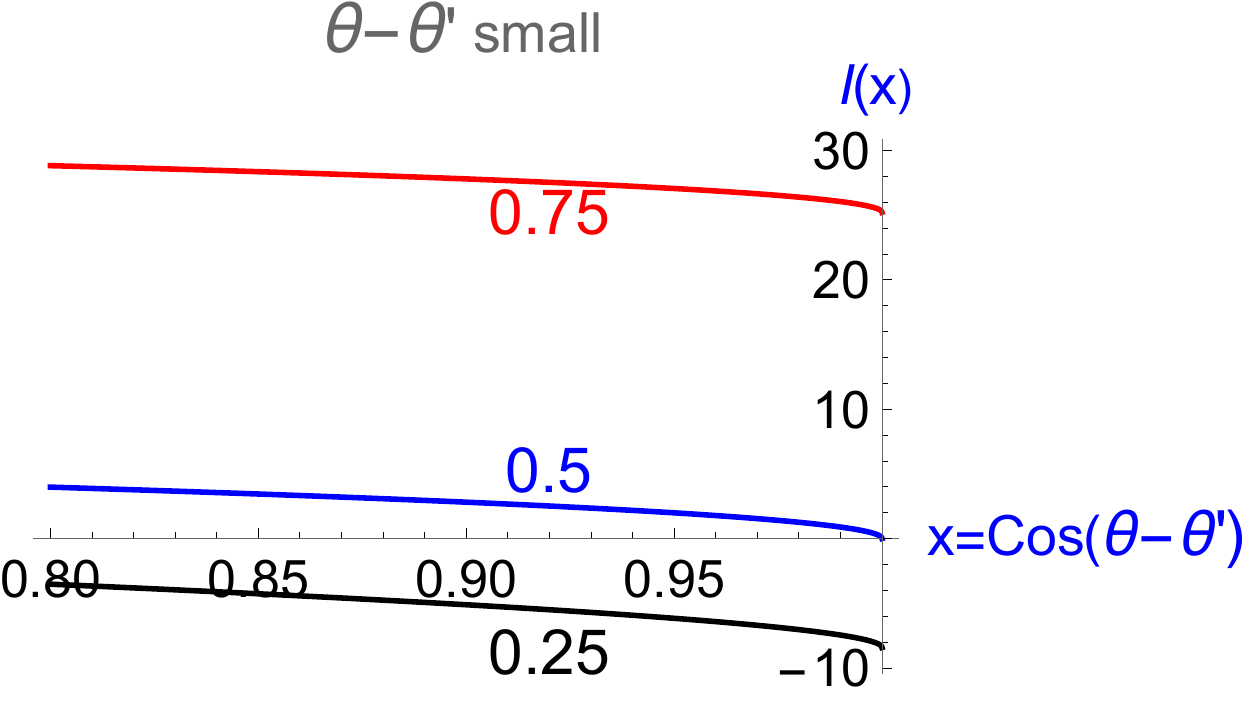}
  }
\caption{ 
(a) Functions $\hat{\mathcal{I}}(\xi)$ versus $\xi= \cos(\theta-\theta') $  for $\alpha=1/2$, the numerical curve (solid line) is very well fitted  by $8.88577 \sqrt{1 - \xi}$ (dashed line). 
(b) Behavior of $\hat{\mathcal{I}}(\xi)$ versus $\xi= \cos(\theta-\theta') $  for small values of $\theta-\theta'$, and various values of the exponent $\alpha$ written explicitly close to the curves.  In the full domain $-1 \le \xi \le 0$   the coefficients in (\ref{eq:Ixi})   are $a=-8.3, \,b=9.25, \,c=3.6$  for  $\alpha=1/4$, and for $\alpha=3/4$ the coefficients are $a=25.2,\, b=8.4, \,c= -0.8$. 
}
\label{fig:Ixi}
\end{figure}

Setting $\xi=\cos(\theta-\theta') $,  and  $\hat{{\mathcal  I}} (\xi)= {\mathcal I}(\theta -  \theta') $  we observe, see Figs.\ref{fig:Ixi}, that the function $\hat {\mathcal I}(\xi)$  is very well fitted by  a low order polynomials in powers of $\sqrt{1-\xi}$
 \begin{equation}  
\hat {\mathcal I}(\xi)=a_{\alpha} +b_{\alpha} \sqrt{1-\xi} + c_{\alpha} (1-\xi)
  \label{eq:Ixi}
\end{equation} 
 or ${\mathcal I}(\theta - \theta')=a_{\alpha} +b_{\alpha}  \vert \theta - \theta' \vert + .. $    where the numerical coefficients $a_{\alpha}, b_{\alpha}..$  depend on the exponent $\alpha$,  see captions. The 
  curves in Fig.\ref{fig:Ixi}- b displays the behavior of $  \hat {\mathcal I}(\xi)$ for  $\xi$ close to unity, namely for small values of $\theta-\theta'$, and for $3$ different values of $\alpha$. 

The problem of computing the average properties of the turbulent mixing layer has been reduced to the search of solutions of the equation 
  \begin{equation}  
  C + D = 0
  \label{eq:C+D}
\end{equation} 
where $ C$ and $ D $ are functions of $g(\theta)$ and its derivatives given in eq.(\ref{eq:sigfinal1}) and eq.(\ref{eq:sigfinal 2}). Those equations have an interesting and non trivial structure reflecting the  fundamental principles they have been derived from. Of course they do not include at all explicit dependence on quantities like a length or a time. Moreover they are both quadratic with respect to $g(.)$, as a  consequence of the fact that no velocity scale should be introduced besides the one arising from the average velocity itself.Equation (\ref{eq:C+D}) has an obvious solution  $g = -  U \sin(\theta - \theta_0)$ with arbitrary constants $U$ and $\theta_{0}$, coming from
  \begin{equation}  
g + g'' = 0.
  \label{eq:gg}
\end{equation} 
 However this solution corresponds to a uniform velocity of strength $U$ in a direction depending on the arbitrary angle   $\theta_0$, that does not correspond to the case of the mixing layer treated here. The problem of the mixing layer corresponds to a solution of equation (\ref{eq:C+D})  with a boundary condition for $g(\theta)$ deduced from the condition that  $u \to  U_{1} $  and $U_{2}$  respectively as $\theta$  tends to $ \pm\pi$ ( $U_{1,2}$ being the incident velocities above and below the board, see Fig.\ref{fig:schema}),   and $v\to 0$ because the  incident flow is parallel to the board. Were those two values $U_{1,2}$ the same, the solution is just  $g = - U \sin \theta$, namely a uniform flow on both sides of the splitter. Below we shall look at the case where the two different values $U_1$ and $U_2$ are close to each other. Because of the nonlinear character of eq.(\ref{eq:C+D}) this makes already a non trivial question.

\subsubsection{ Pressure difference on the two sides of the plate}

Let us return to the general relation (\ref{eq:sig3p}) for the difference of pressure  between the upper and lower parts of the plate. Equation (\ref{eq:sigg}) without the pressure term is 
  \begin{equation}  
  \tilde{  \Sigma}_{ij}=  \rho\, U_{i}U_{j} +  \tilde \sigma_{ij}^{Re}( {\bold{X}})
   \label{eq:tildeSig}
\end{equation} 
 We show in appendix C that  the derivative of the pressure with respect to the angle $\theta$, is given by the integral 
  \begin{equation}    
(p_{t})_{, \theta} (\theta) =  -\rho \, \tilde \gamma \int_{-\pi}^{\pi }  {\mathrm{d}} \theta'    \frac{{\mathrm{d}}}{{\mathrm{d}}\theta} \left(\vert (g + g'')(\theta) \vert^{1-\alpha} {\mathrm{d}} \theta' {\mathcal I}(\theta -  \theta') \right)\vert (g + g'')(\theta') \vert^\alpha \sin 2(\theta-\theta') 
  \label{eq:dp}
\end{equation} 
Looking at the definition of $  \tilde{  \Sigma}_{ij}$ in (\ref{eq:tildeSig}) , we emphasize that the gradient of the total pressure is caused by the effect of the stress tensor, $\tilde \sigma_{ij}^{Re}$,   the first term $U_{i}U_{j}$ yielding a null contribution to  the relation (\ref{eq:ptheta}) leading to the expression of  $(p_{t})_{,\theta}$ in (\ref{eq:dp}).  Integrating by part equation (\ref{eq:dp}),  we get the following relation for the pressure difference between the two sides  of the board
  \begin{equation}    
p_{t}(\pi) -p_{t}(-\pi) =  2\rho\, \tilde\gamma  \iint_{-\pi}^{\pi }  {\mathrm{d}} \theta   {\mathrm{d}} \theta'   \vert (g + g'')(\theta) \vert^{1-\alpha} {\mathcal I}(\theta -  \theta')  \vert (g + g'')(\theta') \vert^\alpha \, (g + g'')(\theta')\, \cos 4(\theta-\theta').
  \label{eq:sautp}
\end{equation}

 \section{Limit of small velocity difference }
 \label{sec:Small}

 \subsection{Scaling law between $\delta \theta$ and $\eta$}  
 \label{sec:scaling}
 From the way the turbulent mixing layer is described, there is one dimensionless number in the data, the relative velocity difference 
   \begin{equation}  
\eta =\frac{U_1 - U_2}{U_1 + U_2}.
    \label{eq:eta}
\end{equation} 
Therefore, besides scaling parameters like the velocity $U = (U_1 + U_2)/2$, every observable quantity of the mixing layer depends on the ratio $\frac{U_1 - U_2}{U_1 + U_2}$, especially the angle 
$ \delta\theta$  
 of the turbulent wedge  which is the most obvious function  to look at. This angle is a function of the dimensionless velocity difference $\eta$, and this
dependence would provide a way of testing theories. 

 After looking at various possibilities for the relationship between $\eta$ and $\delta\theta$ in the limit where both quantities are small, we found only one way to derive such a relation. It is based upon the fact  that the balance of stress depends on the unknown function $g(\theta)$ only. Moreover $g$ appears both in $  C$ and in $ D$ through the combination $(g + g'')$, except the prefactor $g$ in $ C $ which is crucial for finding the relation between $\delta \theta$ and $\eta$.  All the analysis relies on the behavior of  $u$, $v$ and $g$ for small  values of  the two small parameters $\delta \theta$ and $\eta$. 
 
 In the present subsection we derive an estimate of the scaling relation between $\delta \theta$ and $\eta$, leaving the quantitative study to the next subsection. Let us  expand the functions $u$, $v$ and $g$ in powers of $\eta$ in the form $f(\theta)= f^{(0)} + \eta f_{1}+...$. 
 
  Assume first that  $\eta = 0$ which is the case of a uniform velocity  flow incident on the board in the $x$-direction. In this case there is no turbulent flow behind the board, and  the zero order solution is
  \begin{equation}
 \left
 \{ \begin{array}{l}
 u^{(0)} =U \qquad  \text{and}    \qquad v^{(0)}=0\\
  \\
 g^{(0)}(\theta) = -U \sin\theta    \qquad (g+g'')^{(0)}=0  
  \mathrm{.}
\end{array}
\right.
 \label{eq:geta}
\end{equation}
 
 For small $\theta$  the approximation $ g^{(0)}\approx -U \theta$ has to  be used  with  caution because we deal with  expressions having derivatives 
 with respect to $\theta$ which are expected to change rapidly
   in a small interval of width  $\delta\theta$. This fast dependence is linked to the need to extrapolate the velocity field from its value $U_1$ one one side of the mixing layer to $U_2$ on the other side. The corresponding correction to $u(.)$  of order $\eta$ is
   \begin{equation} 
    u(\theta) = U(1+ \eta  u_1(\theta)) + ...
    \label{eq:ueta}
\end{equation} 
where  $u_{1}$ is of order unity because the velocity $u(\theta)$ is equal to  $U$ for $\theta=0$ and equal to $(U \pm \eta)$ for $\theta =\pm \pi$. Similarly we can set
  \begin{equation}
  g=g^{(0)}  +\eta g_{1}
 \label{eq:geta2}
\end{equation} 
 having in mind that 
$g_{1}$ is not necessarily of order unity, because  the terms $g_{1}, g'_{1}$ are linked to $u_{1}$  by the relation
  \begin{equation}
  g_{1}\sin \theta + g'_{1} \cos \theta = - U u_{1}.
 \label{eq:u1g1}
\end{equation} 
Equation (\ref{eq:u1g1}) can be approximated by taking into account  the small angular thickness $\delta \theta$  of the mixing layer which 
makes  the successive  derivatives of $g_{1}$   bigger and bigger, more precisely we have   
  $g'_{1} \sim g_{1}/\delta \theta $ ,  $g''_1 \sim g_1/(\delta \theta)^2$. Therefore (\ref{eq:u1g1})  gives  $g'_{1} \sim  U$, or   
 \begin{equation}  
 g_{1} \, \sim \, U \delta \theta 
    \label{eq:g1}
\end{equation} 
which proves that $g_{1} $ is not of order unity, as announced above. In summary at  first  order with respect to $\eta$ one has 
  $(g + g'' )_{(1)}   \approx \eta  g''_1 \sim \eta g_1/(\delta \theta)^2  $ which 
 becomes   $(g + g'' )_{(1)}   \sim U /\delta \theta$ when using the relation (\ref{eq:g1}),   or 
 \begin{equation} 
 ( g + g'' )   \sim  \eta U/\delta \theta. 
      \label{eq:gsec}
\end{equation} 

Taking this  order of magnitude of $(g + g'')$ in $ D$,  giving to $\theta$ the order of magnitude $\delta \theta$ , and assuming that $\mathcal{I} (\theta-\theta')$ is constant in the small wedge,  and non null, one can estimate $ D \sim  (g'')^{2}$. Consider now $ C$, which has the magnitude $g (g + g'')\sim g^{(0)}\,(g+g'') $.  The relation $ C+D = 0$ leads naturally to the constraint that, if $ C$ and $ D$ are of the same order of magnitude, then   $g^{(0)} \sim g''$ or using (\ref{eq:geta}) and (\ref{eq:gsec}),
  \begin{equation} 
  \delta \theta \sim \eta^{1/2} .
     \label{eq:scalingangle}
\end{equation} 
Note that in the  peculiar case of the exposant $\alpha =1/2$   (and close to this value),  discussed in appendix D,  the above scaling is not valid because $\mathcal{I} (0)=0$ as shown in Fig.(\ref{fig:Ixi})-a. Instead of this approximation, we have to consider  the solution $\mathcal{I} (\theta-\theta')= b \vert (\theta-\theta') \vert $    written in the caption of this figure, which is  of order $ b \delta \theta$, and  the condition $ C+D$ leads to the linear relation $\delta \theta \sim \eta $.

The estimate in (\ref{eq:scalingangle}) is interesting because it shows a kind of amplification of the fluctuations, at least in this limit $ \eta$ small. As far as the order of magnitude is concerned, $\eta$ can be seen as a dimensionless measurement of the given velocity difference driving the instability.  It is quite natural to compare it to the amplitude of the fluctuations of velocity taking place inside  the turbulent wedge. As shown in Sec.\ref{sec:Schwarz}, the variance of the velocity fluctuations
$< u'^2>$ is of order $ \eta^{3/2}  U^2$,  see (\ref{eq:variance}),
 which is much larger (as $ \eta$ tends zero) than the square of velocity difference across the mixing layer, $(U_{1}-U_{2})^{2}$, of order $\eta^{2} U^{2}$. 
 
Another point of interest is the extension of the estimate of the angular width of the turbulent wedge to other situations. We already noticed that such wedges should appear when a parallel flow hits a half plane at an angle with respect to its direction. Applying the same idea as above to this situation one can find the order of magnitude of the angle of the turbulent wedge in the limit of a large Reynolds number. In this limit the perturbation (similar to $\eta$ above) brought by the half-plane is the angle $\beta$ of the half plane with respect of the incoming flow. Let assume that this angle is small. Because it enters in the boundary conditions in the equations for the function $g(\theta)$ like the boundary condition on the two sides of the splitter plate, we could conjecture that the relationship between $\beta$ and $\delta \theta$ displays the same power law as the one between $\beta$ and $\delta \theta$, namely 
\begin{equation}
 \delta \theta \sim \beta^{1/2}, 
     \label{eq:scaling}
\end{equation} 
which also involves  geometrical quantities only.

 \subsection{Solution for small $\eta$ and $\delta \theta$}
 \label{subsec:g-eta-small}
 
 Here we  go further than scaling relations,  by deriving the solution for the time average velocity field for small values of $\eta$ and $\delta \theta$, that allows to give a quantitative expression to (\ref{eq:scalingangle}).
  As usual in this kind of analysis, once the relationship between the various quantities is found, one can get a parameterless equation to be satisfied by the unknown function. In the present case it amounts to find the equation for $u_1$  and $g_{1}$  defined in (\ref{eq:ueta})  and (\ref{eq:u1g1}) as a function of the  angle
     \begin{equation} 
\tilde \theta =\theta / \delta \theta 
    \label{eq:tildtet}
\end{equation} 
where   $\delta \theta $ is  positive and  linked to $\eta$  and should agree ultimately with  the relation (\ref{eq:scalingangle}).  In the final stage of our derivation the small angle  $\delta \theta$ will be defined as the half width at half height  of the velocity derivative $u'(\theta) $, see Fig.\ref{fig:uprim} in Appendix \ref{sec:appsolG}.
In order to handle  functions  of $\tilde \theta$  which are of order unity,  we define $\tilde g_{1}(\tilde \theta)$   and its derivative with respect to $\tilde \theta$  by the relations
\begin{equation}
 g_{1} (\theta)= U \delta \theta  \; \tilde g_{1} ( \tilde \theta)   \qquad
  g'_{1} (\theta)=U \tilde g'_{1} (\tilde \theta) \qquad g''_{1}(\theta) = \frac{U}{\delta \theta} \tilde g''_{1}(\tilde \theta).
%\mathrm{,}
 \label{eq:tildeg1}
\end{equation}
The derivation of the solution of the  equation  for $ g''_{1} (\theta)$  is detailed in Appendix  \ref{sec:appsolG}. This equation  for $ g''_{1} (\theta)$  is deduced   from $C+ D=0$ written in terms of the tilde quantities which becomes
    \begin{equation} 
 \tilde \theta  \tilde g''_1  =  \tilde \gamma {\mathcal I}(0)  \tilde{\mathcal J} \frac{1-\alpha}{\alpha} \vert \tilde g''_1( \tilde \theta) \vert^{-\alpha}  \frac{{\mathrm{d}}} {{\mathrm{d}}\tilde \theta } \vert \tilde g''_1( \tilde \theta) \vert 
   \label{eq:der2g}
\end{equation} 
The solution of (\ref{eq:der2g}) is of the form
 \begin{equation}   
\vert  \tilde g''_{1} (\tilde \theta) \vert = G_{0} \left(1+ (\frac{\tilde \theta }{\tilde \theta_{c}}) ^{2} \right ) ^{-1/\alpha}
     \label{eq:solG}
\end{equation} 
where   $\tilde \theta_{c}$ is a number which depends on  the value of the exponent $\alpha$, see (\ref{eq:tetac})  and $G_{0}=  \,-\,  \tilde g''_{1} (0)$ is positive and depends on the solution itself. Indeed 
we point out that in order to solve  (\ref{eq:der2g}) one has to solve the bootstrap condition that  $\tilde {\mathcal J}$ is given by the value figuring in the solution, and to take into account the behavior of the solution at the boundaries. This procedure allows to get the following quantitative relation between the two small parameters $\delta \theta$ and $\eta$
\begin{equation}   
\delta \theta ^{2}=  \eta \,\left( \tilde \gamma {\mathcal I}(0)\, \frac{4}{\tilde \theta_{c}^{2} }\frac{1-\alpha}{\alpha} \frac{c_{\alpha}} {d_{\alpha}}\right)
     \label{eq:finalscale}
\end{equation} 
 where 
all coefficients in the parenthesis $\tilde \theta_{c} $ , $c_{\alpha}$, $d_{\alpha}$  and  $ {\mathcal I} (0) $ are numerical ones and dimensionless.  
Note that  the coefficient $\tilde \gamma$  in front of the integral defining $\tilde \sigma_{ij}^{Re} $ remains the only one which stays arbitrary. It has to be of  opposite sign with respect to $ {\mathcal I} (0) $, namely must be negative for $\alpha < 1/2$ and positive for $\alpha > 1/2$, as illustrated  in Fig.\ref{fig:I0-coeff}-(a) plotting $ {\mathcal I} (0) $ versus $\alpha$.  The  free parameter $\tilde \gamma$ can be fitted with experiments. Unfortunately few experiments have been made in the regime of small  $\eta$, namely with two incident flows of  quasi equal velocities $U_{1} \approx U_{2}$. However some experimental measures of the turbulent wedge versus $\eta$ are summarized in  Table 3.2 of \cite{these-Rennes}, especially the one of the author which covers the  range  $0.05 \le \eta \le 0.6$. 
 Close to the origin  the experiment displays a non linear regime, see Fig.\ref{fig:Rennes}-(b)  postponed in appendix  \ref{sec:Rennes}, which agrees with our prediction $\delta \theta \sim\eta ^{1/2}$. 
 From this curve, the  value of $\delta \theta / \eta^{1/2}$ allows to fix  the value of the free parameter $\tilde \gamma$ which depends on $\alpha$ as illustrated in Fig.\ref{fig:tildegam}.

\begin{figure}
\centerline{ 
(a)\includegraphics[height=1.5in]{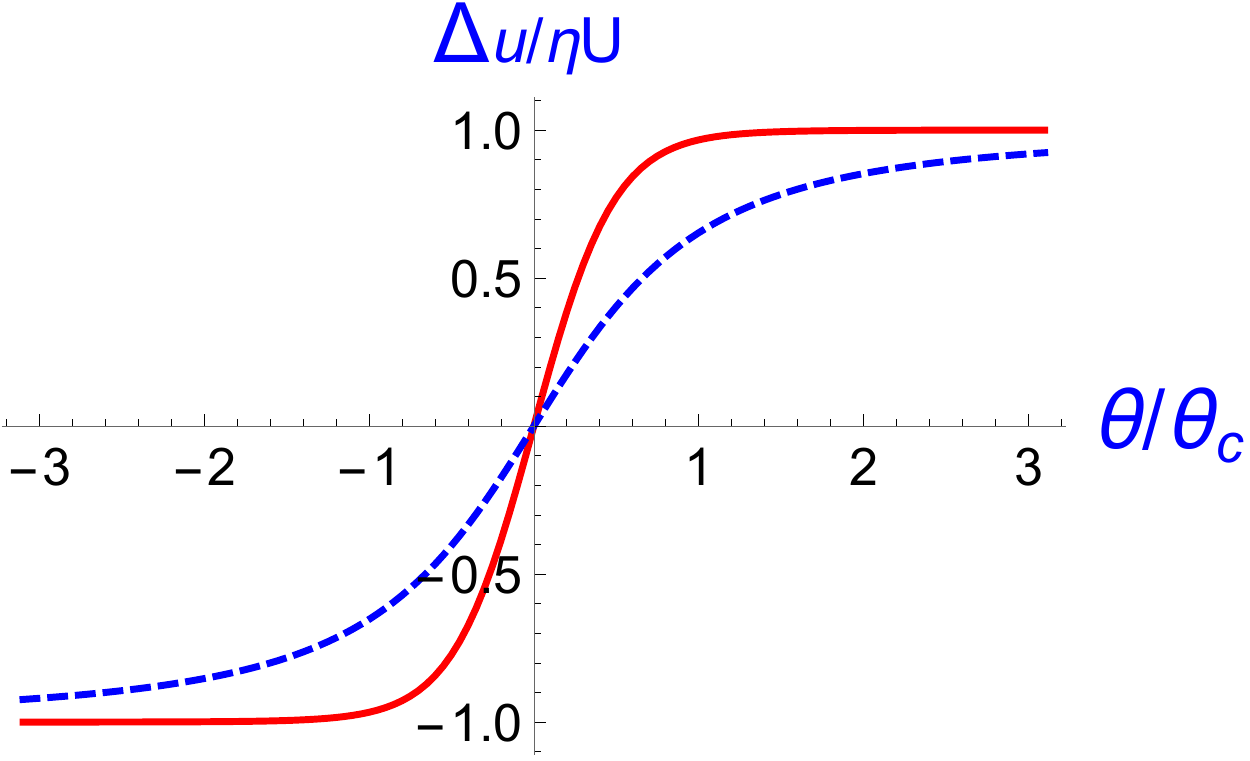}
(b)\includegraphics[height=1.5in]{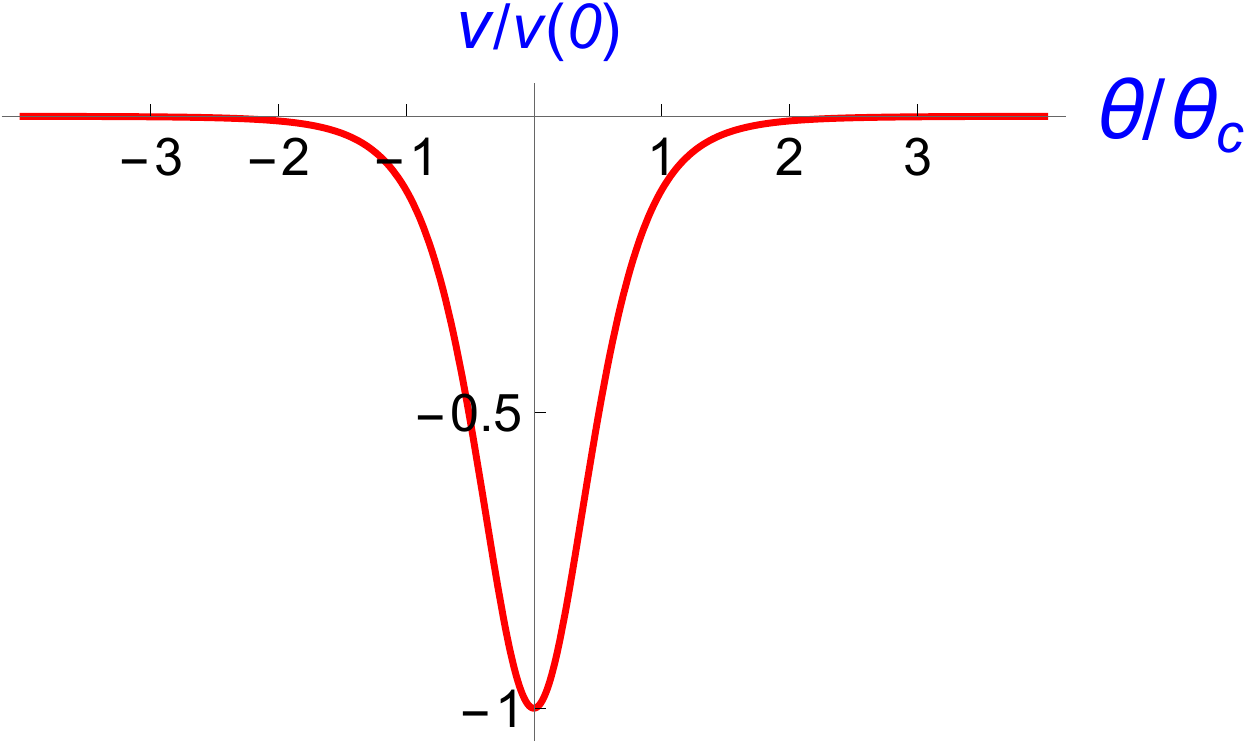}
  }
\caption{(a)  Profile of the x-component of the velocity versus $\theta/\theta_{c}$, equation (\ref{eq:profu}),  for two values of the exponent, $\alpha=0.25$ for the solid red line, $\alpha=0.75$ for the dashed blue line. The increment  $\Delta u= u-U$,  is scaled to $\eta U=(U_{1}-U_{2})/2$  and the polar angle $  \theta$  is scaled to $ \theta_{c}=\delta \theta \; \tilde \theta_{c}$ with $\tilde\theta_{c}$ given in (\ref{eq:tetac1}). The asymptotic values of  $u$ are $U_{1}$ (resp. $U_{2}$ ) as $\theta$  tends to $\pm \infty$.  
(b) Profile of the y-component of the velocity, equation (\ref{eq:profv}),   versus $\theta /\theta_{c}$  for $\alpha=0.25$,  $v$ is scaled to  $v(0)= (1/2)\eta^{3/2} k^{1/2}(U/d_{\alpha}) \alpha/(1-\alpha)$.}
\label{fig:profil}
\end{figure}

Finally the  profile of the longitudinal and transverse velocity components, deduced from the solution (\ref{eq:solG})  are given by the expressions
 \begin{equation} 
u(\theta) -U = \frac{\eta U}{d_{\alpha}} \int _0^{\theta/\theta_{c}}  {\mathrm{d}}y \, (1+y^{2})^{-1/\alpha}
    \label{eq:profu}
\end{equation}   
where $d_{\alpha}$ is the value of the integral for $\theta/\theta_{c}=\infty$, given in (\ref{eq:dalpha}),
and 
 \begin{equation} 
v(\theta) = \,-\,  \frac{ \eta^{3/2}U k^{1/2} \tilde \theta_{c}}{d_{\alpha}}\,\frac{\alpha}{2(1-\alpha)}\left (1+\frac{\theta^{2}}{\theta_{c}^{2}}\right )^{-\frac{1-\alpha} {\alpha}},
    \label{eq:profv}
\end{equation}  
where $\theta_{c}=  \tilde \theta_{c} \delta \theta $  and $k$ is the ratio $\delta \theta^{2}/\eta$ in (\ref{eq:finalscale})
 \begin{equation} 
  k=\tilde \gamma {\mathcal I}(0)\, \frac{4}{\tilde \theta_{c}^{2} }\frac{1-\alpha}{\alpha} \frac{c_{\alpha}} {d_{\alpha}}
    \label{eq:k}
\end{equation}  
 is a positive constant  because the product $\tilde \gamma {\mathcal I}(0)$ has to be positive. The profiles are drawn in Fig.\ref{fig:profil}.  In (a) the role of the exponent $\alpha$ appears artificially  because of the scaled abscissa, although the two curves have the same halfwidth $\delta \theta$ by definition.

\subsubsection{pressure difference on the plates} 
Let us  finally consider the order of magnitude of the pressure difference between the two plates. As shown in Appendix C, we get
   \begin{equation}   
 p(\pi)-p(-\pi) \sim (\eta U)^{2}
     \label{eq:appC9}
\end{equation} 
Therefore the  difference of pressure reflects the lift force which is quadratic with respect to $(U_{2}-U_{1}) $, as expected.

 \subsubsection{ order of magnitude of   $\tilde \sigma_{ij}^{Re} $ and $\sigma_{ii}^{Re,p} $ }
\label{sec:Schwarz}
As detailed  in the appendix, the order of magnitude of the components of $\tilde \sigma_{ij}^{Re} $ are
\begin{equation}
 \left \{ \begin{array}{l}
\tilde \sigma_{xx}^{Re}  = - \tilde \sigma_{yy}^{Re} \,\sim \eta^{5/2}\\
  \\
 \tilde  \sigma_{xy}^{Re}  \, \sim  \eta^{3/2}
  \mathrm{.}
\end{array}
\right. \label{corsigI}
\end{equation}
 As expected these relations do not satisfy the realizability conditions (\ref{eq:appDcondreal}),  because  one diagonal components  is negative, moreover $ \sigma_{xy} ^{2} \; > \;  \sigma_{xx} \,\sigma_{yy}  $ which is
inconsistent for correlation functions.

Let us now consider the order of magnitude of the diagonal elements  of the tensor $\sigma_{ij}^{Re,p} $  defined in (\ref{eq:sig22}). In the mixing layer case they become
 \begin{equation} 
 \sigma_{ii}^{Re,p}(\theta) =   \rho \, \tilde \gamma \vert (g + g'') \vert^{1-\alpha} \int {\mathrm{d}} \theta' {\mathcal I}(\theta -  \theta')  \vert (g + g'')(\theta') \vert^{\alpha}  (g + g'')(\theta').
   \label{eq:sig-p}
\end{equation}   
They are of order $(g'')^2 $ times $\delta \theta$ ( which comes from the integration), that gives 
    \begin{equation}   
 \sigma_{ii}^{(Re,p)} \sim \eta ^{3/2},
     \label{eq:variance}
\end{equation} 
which is of the same order as $ \tilde  \sigma_{xy}^{Re} $.  Therefore  the realizability conditions (\ref{eq:appDcondreal}) can be fulfilled by the sum of the two tensors in (\ref{eq:sigRe1}).
Moreover the  off diagonal element  $ \tilde  \sigma_{xy}^{Re} $ contains an integrand smaller than the  diagonal one $\sigma_{ii}^{(Re,p)}$ because of the presence of $\cos 2(\theta-\theta')$ in $ \tilde  \sigma_{xy}^{Re} $  compared to unity in $\sigma_{ii}^{(Re,p)}$. It follows that   the realizability conditions are fulfilled  by taking  the same constant for the two tensors in (\ref{eq:sig11})-(\ref{eq:sig22}), namely $\gamma= \tilde \gamma$, that correspond to on-axis fluctuations $<u'v'>, <u'^{2}>,<v'^{2}>$ of quasi-equal amplitude for small $\eta$ within the frame of our model.   Actually the ratio $\gamma/\tilde \gamma$   is generally larger than unity and has to be adjusted with experimental results.We haven't found any profile  of velocity fluctuations for small values of $\eta$, most of them being concerned  by ratios  $U_{1}/U_{2} $ of order few units.
 
%However we have found The experimental work  of K. Sodjavi \cite{these-Rennes} shows that this ratio is close to $2$ for $\eta=0.33$ (see  the profiles of $<u'v'>, <u'^{2}>$ and $<v'^{2}>$ in Fig. 3.4 of Ref. \cite{these-Rennes}). 
% In the literature  we found few experimental works  devoted to  small values of $\eta$,  most of them being concerned  by ratios  $U_{1}/U_{2} $ of order few units. 
 Nevertheless we  found an experimental study extending from $\eta=0.05$ up to $\eta=0.6$ \cite{these-Rennes}, with many references to other works.  In Table 3.2  of  \cite{these-Rennes} the author compares his measurement  of   the turbulent  wedge angle versus $\eta$ with  other measurements.  Only two experiments cover the range of small $\eta$ values, the one of the author and the one of  Mehta \cite{mehta}.
 In this domain  both curves $\delta \theta$ versus $\eta$,  display similar non linear behavior [see Fig.\ref{fig:Rennes}-(a)]  which agrees  with our prediction $\delta \theta \sim \eta^{1/2}$ [see Fig.\ref{fig:Rennes}-(b)], as detailed in Appendix \ref{sec:Rennes}.
%  although the author concludes that the width of the turbulent domain grows linearly with $\eta$ in agreement with the Boussinesq model.
Those data allow to give a numerical value to the free parameter $\gamma$ (then also $ \tilde \gamma$, at least approximately)  for a given value of the exponent $\alpha$, the only parameter which remains free at this stage, see Fig.\ref{fig:tildegam}. Note that $\alpha$ could depend on the experimental set up.

 \section{Conclusions and perspectives}
  \label{sec:Concl}
  
  In this paper we wrote fully  explicitly the integral equation for the balance of momentum including the closure of the turbulent stress introduced in ref  \cite{chaos} on the basic assumption that dissipation  is caused by singular events described by solutions of Euler's equation. Because of the fully explicit character of this closure it is possible to obtain results more detailed than what was derived long ago by Prandtl and Landau.  Notice that this closure yields equations with all the expected scaling laws, which is not surprising because the equations are derived to satisfy those scaling laws, but also that more quantitative properties can be put in evidence, including the effect of boundary conditions. This point is non trivial because the boundary conditions make often a non trivial issue for integral equations. 
  
 Our detailed analysis  applied to the turbulence behind the plate in mixing layer set up is performed in the limit of a small velocity difference $\eta$. In this limit we have found only  few experiments reported, nevertheless they  agree with our prediction that the angular spreading of the turbulent domain scales as $\eta^{1/2}$. From this agreement one is able to extract the value of  one the  three free parameters $\alpha$, $\tilde \gamma$ and $\gamma$  of our model. The more general case of $\eta$ of order unity is in progress. We hope it would yield a deeper comparison between our model and  numerical or  experimental data. In particular this would allow to give the value, even approximate, of the exponent $\alpha$.    
  
    Of course one can also hope to get solutions of the momentum balance in situations more complex than the one considered here. We think first to an axisymmetric wake like the one behind a disc perpendicular to the incoming flow. This adds one more coordinate, the position in the flow direction, added to the radius in the perpendicular direction, but the question of imposing the boundary condition is nontrivial. We plan to study those flows  in a near future.

\section*{Acknowledgement}
We greatly acknowledge Christophe Josserand and Sergio Rica for fruitful and stimulating discussions.

\appendix

\section{ Contribution of the inertia to equation (\ref{eq:sig3}) }
\label{sec_appA}

Here we derive  equation (\ref{eq:sigfinal1})   which represents the contribution of the inertia  to equation (\ref{eq:sig3}). We have to  insert  in (\ref{eq:sig3})  the first term $U_{i }U_{j}$  of  the tensor
\begin{equation}   
 \tilde \Sigma =  U_{i}U_{j}+ \tilde \sigma_{ij}^{Re}.
     \label{eq:app1}
\end{equation} 
 With $U_{x} =u $ and $U_{y} = v$,   we get
 
 \begin{equation}   
C=  \sin\theta \cos \theta \left((v^2)_{,\theta} - (u^2)_{,\theta} \right) +(\cos^2 \theta -\sin^2 \theta) (u v_{,\theta}+ v u_{, \theta}).
     \label{eq:app2}
\end{equation} 
By definition of $(u,v)$  in polar coordinate
\begin{equation}
 \left \{ \begin{array}{l}
u(\theta) =  - (g \sin \theta  + g' \cos \theta)  \\
v(\theta)  =  g \cos \theta   - g' \sin \theta
\end{array}
\right.
 \label{eq:appAuv}
\end{equation}
we have
\begin{equation}
 \left \{ \begin{array}{l}
u_{,\theta}= -(g+g'') \cos \theta   \\
v_{,\theta}= -(g+g'') \sin \theta
\end{array}
\right.
 \label{eq:deruv}
\end{equation}
Inserting these derivatives in (\ref{eq:app2}) we obtain
 \begin{equation}   
\frac{C}{g+g''} =  \sin 2\theta ( - g \sin 2\theta  - g' \cos 2 \theta)  +  \cos 2\theta ( - g \cos 2\theta  + g' \sin 2 \theta) 
     \label{eq:app4}
\end{equation} 
which reduces to 
 \begin{equation}   
\frac{C}{g+g''} =  - g ,
     \label{eq:app5}
\end{equation} 
or (\ref{eq:sigfinal1})

\section{ Contribution of the  stress tensor $\tilde \sigma_{ij}^{Re}$  to  equation (\ref{eq:sig3}) }
\label{sec_appB}

Here we derive equation (\ref{eq:sigfinal 2}) which represents the contribution of the non diagonal  Reynolds stress tensor $\tilde \sigma_{ij}^{Re}$ to  equation (\ref{eq:sig3}). We have to  insert  in (\ref{eq:sig3})  the second term of  the tensor  
 $\tilde \Sigma $ defined in (\ref{eq:app1}), that gives
 \begin{equation}   
D =  \frac{1}{2} \sin 2 \theta  (\tilde \sigma_{yy}^{Re} - \tilde \sigma_{xx}^{Re})_{,\theta}  +  \cos 2\theta (\tilde \sigma_{xy}^{Re})_{,\theta}.
     \label{eq:appB1}
\end{equation} 
Using  (\ref{eq:tautild}) and   (\ref{eq:tauxx})-(\ref{eq:tauxy}), the components of $\tilde{ \tau}_{ij}  $  become 
\begin{equation}
 \left \{ \begin{array}{l}
\tilde{ \tau}_{xx} = \sin 2\theta (g+g'')= - \tilde{ \tau}_{yy} \\
  \\
\tilde{ \tau}_{xy} = -\cos 2\theta (g+g'')=  \tilde{ \tau}_{yx}   
  \mathrm{.}
\end{array}
\right. \label{eq:tildtauij}
\end{equation}
Inserting these latter expressions in the integrand of (\ref{eq:sigfin}), we get 
 \begin{equation}   
D =  B_{1} \sin 2 \theta    +  B_{2}  \cos 2\theta 
     \label{eq:appB4}
\end{equation} 
where
  \begin{equation}   
B_{1}=  - \tilde \gamma  \frac{{\mathrm{d}}}{{\mathrm{d}}\theta} \left(\vert (g + g'')(\theta) \vert^{1-\alpha} \int_{-\pi}^{\pi}  {\mathrm{d}} \theta' {\mathcal I}(\theta -  \theta') \right) \vert (g + g'')(\theta') \vert^\alpha\, (g + g'')(\theta') \,\sin 2\theta'
     \label{eq:appB5}
\end{equation} 
and
  \begin{equation}   
B_{2}=  -  \tilde \gamma  \frac{{\mathrm{d}}}{{\mathrm{d}}\theta} \left(\vert (g + g'')(\theta) \vert^{1-\alpha} \int_{-\pi}^{\pi}  {\mathrm{d}} \theta' {\mathcal I}(\theta -  \theta') \right) \vert (g + g'')(\theta') \vert^\alpha\,(g + g'')(\theta')\,\cos 2\theta'
     \label{eq:appB6}
\end{equation} 
Equations (\ref{eq:appB4}), (\ref{eq:appB5}), (\ref{eq:appB6}) are equivalent to the compact form  (\ref{eq:sigfinal 2}).

\section{Pressure difference between the two sides of the plate }
\label{sec_appC}

Here we derive the expression (\ref{eq:dp})  for the pressure gradient and  the  pressure  difference (\ref{eq:sautp}) between the two sides of the plate. Looking at  equation 
 \begin{equation} 
   ( p_{t})_{,\theta}= - \sin^2 \theta \; \tilde{\Sigma}_{xx, \theta} -  \cos^2 \theta\;  \tilde{\Sigma}_{yy, \theta} + 2 \sin \theta  \cos \theta \; \tilde{\Sigma}_{xy, \theta} 
    \label{eq:appC1}
\end{equation} 
already written in (\ref{eq:ptheta}),
with $ \tilde \Sigma$ defined in (\ref{eq:app1}),  let first consider the contribution of  the inertial term $U_{i}U_{j}$   to this expression for $p_{,\theta}$. This contribution is the sum $c_{1}+ c_{2}$ with
 \begin{equation}   
c_{1}= -  2  u u_{,\theta} \sin^2 \theta  - 2 v v_{,\theta}  \cos^2 \theta 
     \label{eq:appC1}
\end{equation} 
and
 \begin{equation}   
c_{2}=  2 ( v\, u_{,\theta} \,+ \, u \, v_{,\theta})  \sin \theta   \cos \theta 
     \label{eq:appC2}
\end{equation} 
Using  (\ref{eq:u})-(\ref{eq:v}) and (\ref{eq:deruv}),  we get the relations
 \begin{equation}   
c_{1}/ (g+g'')=   g \sin 2\theta \cos 2\theta - g' (\sin 2 \theta )^2 = \, - \,c_{2}/(g+g'')
     \label{eq:appC3}
\end{equation} 
which shows that the inertial term does note contribute to the gradient of pressure in such 2D geometry, as written in the text.

Consider now the effect of the  tensor $ \tilde \sigma_{ij}^{Re}$,  defined in (\ref{eq:sig11}). 
Separating, as above, the diagonal and non-diagonal terms of this  tensor, gives
 \begin{equation}   
(p_{t})_{,\theta}= d_{1}  + d_{2},
     \label{eq:appCo}
\end{equation}  
 with 
 \begin{equation}   
d_{1}=   -(\tilde  \sigma_{xx}^{Re})_{,\theta} \sin^2 \theta  - (\tilde \sigma_{yy}^{Re})_{,\theta}  \cos^2 \theta =  \cos 2\theta  (\tilde \sigma_{xx}^{Re})_{,\theta} 
     \label{eq:appC4}
\end{equation}  because $\tau_{xx}=-\tau_{yy}$,
and
 \begin{equation}   
d_{2}=   \sin 2 \theta   (\tilde \sigma_{xy}^{Re})_{,\theta} 
     \label{eq:appC5}
\end{equation} 
Inserting in these relations the equations (\ref{eq:tauxx}) to (\ref{eq:tautild}), we get 
 \begin{equation}   
d_{1}=  \tilde  \gamma \rho  \cos 2\theta  \frac{{\mathrm{d}}}{{\mathrm{d}}\theta} [  \vert (g + g'') \vert^{1-\alpha} \int {\mathrm{d}} \theta' {\mathcal I}(\theta -  \theta') ] \,\vert (g + g'')(\theta') \vert^{\alpha}  (g + g'')(\theta')  \sin (2 \theta'), 
     \label{eq:appC6}
\end{equation} 
and
 \begin{equation}   
d_{2}=  - \tilde \gamma \rho  \sin 2\theta  \frac{{\mathrm{d}}}{{\mathrm{d}}\theta}[ \vert (g + g'') \vert^{1-\alpha} \int {\mathrm{d}} \theta'   {\mathcal I}(\theta -  \theta')] \, \vert (g + g'')(\theta') \vert^{\alpha}  (g + g'')(\theta')  \cos ( 2\theta'), 
     \label{eq:appC7}
\end{equation} 

In order to get to the pressure difference on the plate,  we have to integrate $d_{1}$ and $d_{2}$ over the variable $\theta$   running from $-\pi$ to $+\pi$. Integrating by parts the two terms  in (\ref{eq:appCo}),
 and taking into account  the condition $g+g'' =0$ at the boundary which  allows to cancel the constant term of this integration (the term depending on the boundary values), we obtain
  \begin{equation}   
 p_{t}(\pi)-p_{t}(-\pi) = 2  \tilde \gamma \rho  \iint_{-\pi} ^{+\pi} {\mathrm{d}} \theta' \;{\mathrm{d}}\theta' \cos 2(\theta-\theta') \;  \vert (g + g'')(\theta) \vert^{1-\alpha}  {\mathcal I}(\theta -  \theta')
\vert (g + g'')(\theta') \vert^{\alpha}  (g + g'')(\theta').
    \label{eq:appC8}
\end{equation} 
 From this expression we can derive  the difference of pressure on the two sides of the plate in the limit of small difference between the two incident velocity. Using the  arguments   developed in  section \ref{sec:Small} , we have $   (g + g'') \sim  \eta U/\delta \theta $  in the turbulent domain  of angular extension $\delta \theta$. It follows  that the order of magnitude of the integrand in (\ref{eq:appC8})  is of order $(\eta U/ \delta \theta)^{2}$, which has to be multiplied by  $(\delta \theta )^{2}$ to represent the order of magnitude of the pressure difference. We obtain  
   \begin{equation}   
p_{t}(\pi)-p_{t}(-\pi) \sim (\eta U)^{2}.
     \label{eq:appC9}
\end{equation} 
We find that  the  lift force is quadratic with respect to 
 $(U_{2}-U_{1}) $, as expected from the Kutta-Jukovsky theorem.

\section{ Derivation of the velocity profil and ratio $\delta \theta ^{2}/ \eta$ }
\label{sec:appsolG}

Here derive the solution of $C+D=0$ for small  values of the two parameters $\eta$  and $\theta$.   Assuming that  $ {\mathcal I}(\theta-\theta')= {\mathcal I}(0)$ and $\cos(\theta-\theta')=1$ in the integrant of (\ref{eq:sigfinal 2}), (see below for the justification), we have to solve 
  \begin{equation} 
- g(\theta) (g+g'')=  \tilde \gamma {\mathcal I}(0) {\mathcal J} \frac{{\mathrm{d}}} {{\mathrm{d}} \theta } \vert (g+ g'')\vert ^{1-\alpha}
   \label{eq:appo}
\end{equation} 
where
 \begin{equation} 
 {\mathcal I} (0) = \frac{2}{1-\alpha} \int _0^{\infty}  {\mathrm{d}} \zeta   \frac {1}  {\zeta^{\alpha} (1-\zeta)}, 
    \label{eq:appo1}
\end{equation}   
 \begin{equation} 
 {\mathcal J}=  \int_{-\infty}^{\infty}  {\mathrm{d}}  \theta' (g+ g'')(\theta')\vert (g+ g'') (\theta') \vert^\alpha. 
    \label{eq:appo2}
\end{equation} 
In a second step, after having found  the solution for $g+g''$, we have to integrate   
  \begin{equation}   
   \left \{ \begin{array}{l}
   u'( \theta)= -(g+g'')\cos \theta\\
   \\
v'( \theta)= -(g+g'')\sin \theta \mathrm{.}
\end{array}
\right.
 \label{eq:upvp}
\end{equation} 
together with the  four boundary conditions
 \begin{equation} 
u(\pm\pi)\,=\, U (1\pm \eta)  \qquad  v(\pm\pi)\,=\, 0
    \label{eq:4cl}
\end{equation}

\subsection{ Solution for  $g^{(0)} \approx -U \theta$ }

For  $\eta=0$,   the velocity components are
  \begin{equation}
 u^{(0)} =U   \qquad  \text{and}  \qquad v^{(0)}=0 
 \label{eq:sol0}
\end{equation}
and  the $g$ function is
  \begin{equation}
g^{(0)}(\theta) = -U \sin \theta    \qquad  \text{or}  \qquad (g+g'')^{(0)}=0 
 \label{eq:sol01}
\end{equation}

\begin{figure}
\centerline{ 
\includegraphics[height=1.5in]{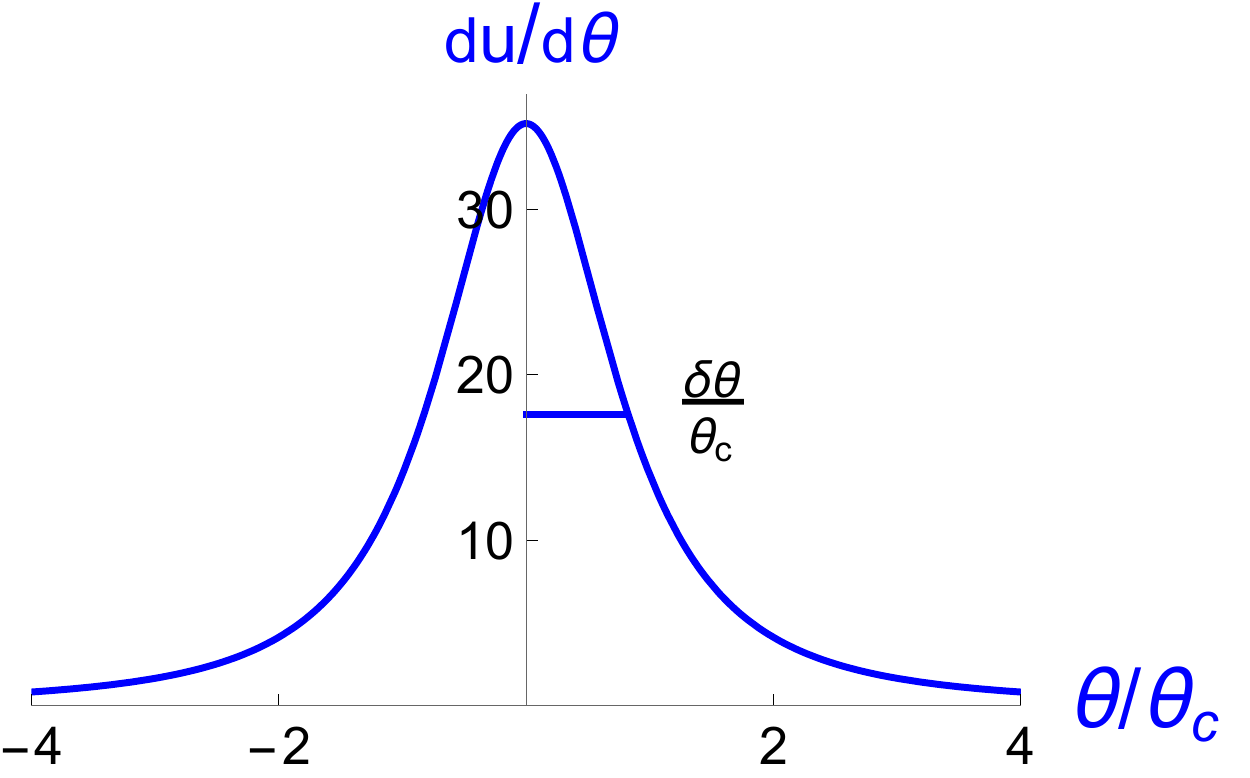}
   }
\caption{ Derivative of the velocity component, $u_{,\theta}$ versus $\theta/\theta_{c}$, for $\eta=0.26$, $\alpha=3/4$. The half-width at half-height is $\delta \theta= 0.02$ ( data of \cite{these-Rennes}) and  (\ref{eq:tetac1}) yields $\theta_{c}=0.025$
 .}
\label{fig:uprim}
\end{figure}

Using these expressions for the  leading order solution  we have to expand all functions in powers of $\eta$ and $\theta$, that will allow to get finally a  scaling relation between the two small parameters $\theta$ and $\eta$.   
Let us define the dimensionless parameter
     \begin{equation} 
\tilde \theta =\theta / \delta \theta 
    \label{eq:tildtet}
\end{equation} 
where   $\delta \theta $  (a positive quantity)  is the small angular aperture of the turbulent domain. More precisely  $\delta \theta $  will be defined  below
as the half width at half height  of the velocity derivative $u'(\theta) $, see Fig.\ref{fig:uprim}.
We can use the relations (\ref{eq:g1})  and  (\ref{eq:gsec}) to define  functions  of $\tilde \theta$  which are of order unity. We set %  we define $\tilde g_{1}(\tilde \theta)$  by the relation
     \begin{equation} 
 g_{1} (\theta)= U \delta \theta  \; \tilde g_{1} ( \tilde \theta)
     \label{eq:tildg1}
\end{equation}  
that leads to
  \begin{equation} 
  g'_{1} (\theta)=U \tilde g'_{1} (\tilde \theta) \qquad g''_{1}(\theta) = \frac{U}{\delta \theta} \tilde g''_{1}(\tilde \theta)
       \label{eq:tildgp1}
\end{equation}  
where  $\tilde g'_{1}, \tilde g''_{1}$, are the  derivatives of $\tilde g_{1}$ with respect to the variable $\tilde \theta$.
  Thanks to those scaling the boundaries in the integral defining $D$ are sent to plus and minus infinity. In terms of the tilde quantities, the left hand side of (\ref{eq:appo}) is
   \begin{equation} 
 C /\rho=\, - \, \eta U ^{2} \tilde \theta  \, \tilde g''_1(\tilde \theta). 
      \label{eq:tildeC}
\end{equation} 
The right hand side 
    \begin{equation}  
 -D/ \rho  =  \tilde \gamma {\mathcal I}(0)\eta^{2} \frac{{\mathrm{d}}}{{\mathrm{d}} \theta } \left(\vert g''_1(\theta) \vert^{1-\alpha}\right) \int_{-\infty}^{\infty}  {\mathrm{d}}  \theta'  g''_1(\theta')\vert  g''_1 ( \theta') \vert^\alpha
   \label{eq:sigfinal 2.1}
\end{equation} 
becomes in tilde variables
    \begin{equation}  
 -D/ \rho  =  \tilde \gamma {\mathcal I}(0)(1-\alpha) \left(\frac{U\eta}{\delta \theta}\right)^{2} \tilde {\mathcal J} \,  \vert  \tilde g''_1( \tilde \theta) \vert^{-\alpha}  \frac{{\mathrm{d}}}{{\mathrm{d}}\tilde \theta } \vert  \tilde g''_1 ( \tilde \theta) \vert
   \label{eq:DI2}
\end{equation} 
where we introduced
  \begin{equation}  
\tilde {\mathcal J} = \int_{-\infty}^{\infty}  {\mathrm{d}} \tilde \theta'   \tilde g''_1(\tilde \theta')\vert  \tilde g''_1 ( \tilde \theta') \vert^\alpha 
   \label{eq: tildeJ}
\end{equation} 
In these expressions the sign of $g''_{1}$  is known because one has 
  \begin{equation}   
u'(\theta)=-\eta \frac{U}{\delta \theta}\tilde g''_{1}
     \label{eq:up2}
\end{equation} 
and  we expect that the slope of the the velocity profile $u(\theta)$ is positive for $\eta >0$  (the case  schematized in Fig.\ref{fig:schema}),  and negative for $\eta <0$, that imposes $\tilde g''_{1} ( \tilde \theta)<0$ in both cases. Setting   $ \tilde g''_1( \tilde \theta)= - \vert \tilde g''_1( \tilde \theta)\vert$,   and  $ \tilde {\mathcal J}  = - \vert  \tilde {\mathcal J} \vert$, equation
 (\ref{eq:appo}) becomes
  \begin{equation}   
 \tilde \theta= \frac{\eta}{\delta \theta ^{2}} \frac{1-\alpha}{\alpha} \tilde \gamma \vert \tilde {\mathcal J} \vert  {\mathcal I}(0) \frac{{\mathrm{d}}}{{\mathrm{d}}\tilde \theta }  \vert  \tilde g''_1( \tilde \theta))\vert^{-\alpha}  
     \label{eq:G1}
\end{equation}
By integration we obtain
 \begin{equation}   
\vert  \tilde g''_{1} (\tilde \theta) \vert = G_{0} \left(1+ (\frac{\tilde \theta }{\tilde \theta_{c}}) ^{2} \right ) ^{-1/\alpha}
     \label{eq:solGapp}
\end{equation} 
where  $G_{0}=  \,-\,  \tilde g''_{1} (0)$ is positive,  and  
 \begin{equation} 
 \tilde \theta_{c} ^{2}= 2 (\eta/\delta \theta^{2})\, ((1-\alpha)/\alpha) \,  \vert \tilde \gamma \,\tilde {\mathcal J}  \,{\mathcal I}(0) \vert \, G_{0}^{-\alpha}.
      \label{eq:tetac3}
\end{equation}
Now we have to take into account that  the solution (\ref{eq:solGapp}) for $ \tilde g''_1$ has to be put in  $\tilde {\mathcal J} $  
defined in (\ref{eq:appo2}).
 We get  $\vert \tilde {\mathcal J} \vert  = (2   \tilde \theta_{c}G_{0} ^{\alpha+1}\, c_{\alpha })$
where 
  \begin{equation}   
c_{\alpha}=   \int_{0}^{\infty} {\mathrm{d}}y \,(1+y^{2})^{- (1+\alpha)/ \alpha} \,=\,  \sqrt \pi \frac{\Gamma (\frac{\alpha+1}{\alpha} -\frac{1}{2})}{ 2\Gamma (\frac{\alpha+1}{\alpha })}, 
     \label{eq:calfa}
\end{equation} 
and $\Gamma(.)$ is the usual Gamma function.
Putting the latter relations in (\ref{eq:tetac3}) gives
 \begin{equation} 
  \tilde \theta_{c} = 4\frac{\eta}{\delta \theta ^{2}}\frac{1-\alpha}{\alpha} \tilde \gamma {\mathcal I}(0) G_{0} c_{\alpha}.
     \label{eq:tetac1}
\end{equation} 
  Equation (\ref{eq:tetac1}) implies that the product $\tilde \gamma{\mathcal I}(0) $ has to be positive, it follows that the factor $\tilde \gamma $ in front of $\tilde \sigma_{ij}^{Re}$ must be positive when the exponent $\alpha$ is bigger than $1/2$, and negative in the opposite case.

Now we have to take into account the boundary conditions of the velocity field which can be written as
\begin{equation}   
\int_{0}^{\infty}  {\mathrm{d}}\theta \, u'(\theta)=\eta U.
     \label{eq:boundary}
\end{equation} 
Using (\ref{eq:up2}) this relation  becomes 
\begin{equation}   
G_{0}  d_{\alpha}\tilde \theta_{c}=1
     \label{eq:bound}
\end{equation} 
where 
\begin{equation}   
d_{\alpha}=    \int _{0}^{\infty}  dy(1+y^{2 })^{-1/\alpha  } = \sqrt{\pi}   \frac {\Gamma(1/\alpha-1/2)} {2\Gamma(1/\alpha)}.
     \label{eq:dalpha}
\end{equation}

\begin{figure}
\centerline{ 
(a) \includegraphics[height=1.5in]{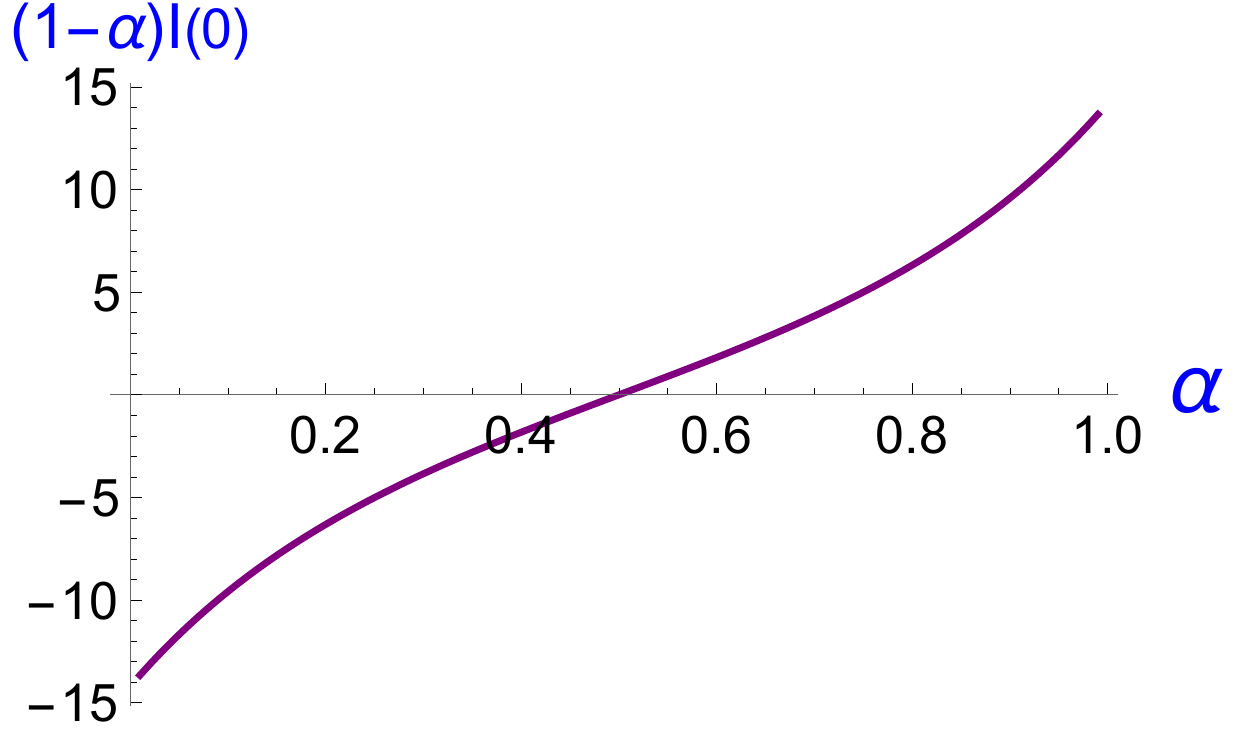}
  (b) \includegraphics[height=1.5in]{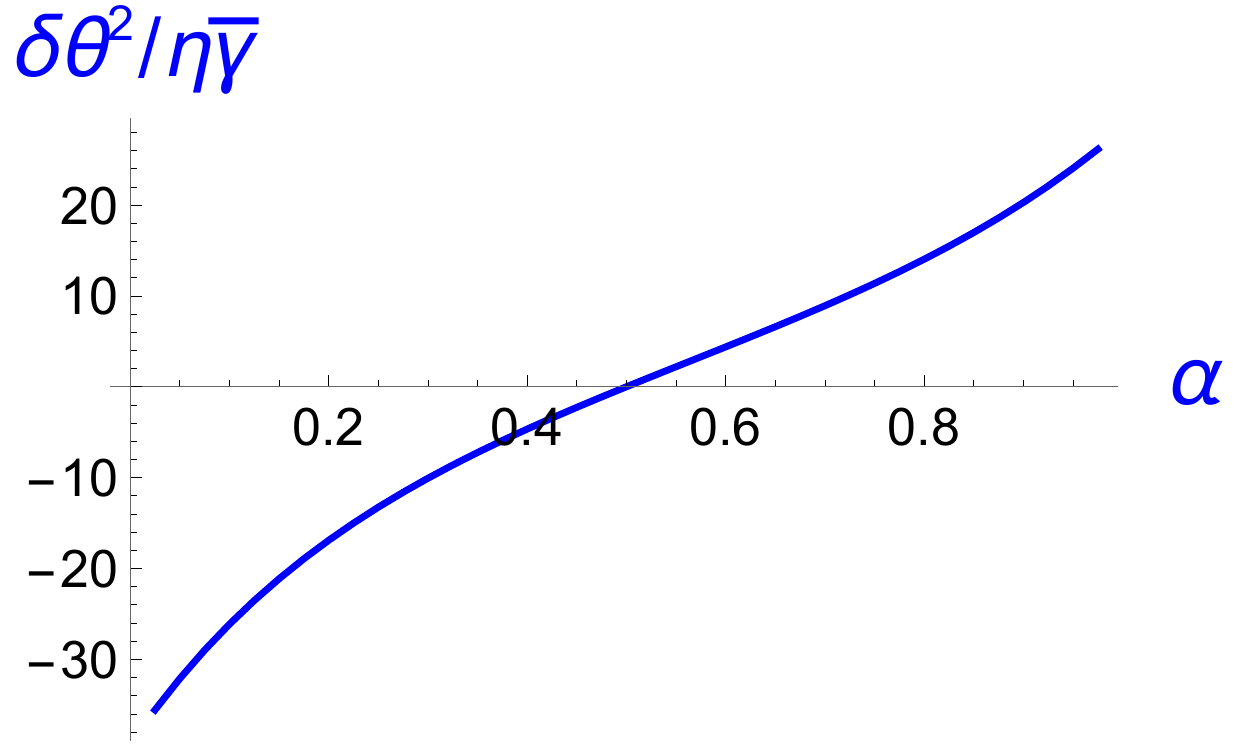}
   }
\caption{
 (a)  $(1-\alpha){\mathcal I} (0)$ versus $\alpha$, see equation (\ref{eq:defI0}). 
 (b) Ratio $\delta \theta ^{2} /  \eta \tilde \gamma$ versus $\alpha$  given in equation (\ref{eq:finalscale}). The product $\tilde \gamma{\mathcal I} (0)$ has to be positive, then $\tilde \gamma <0$ for $\alpha <1 /2$.}
\label{fig:I0-coeff}
\end{figure}

Finally, we point out that  the width of the solution $ g''(\theta) \propto u'(\theta)$   depends on the value of the exponent $\alpha$.  In order to take this dependence into account we can define the angular width of the turbulent wedge  as the half width at half height of  $u'(\theta)$, see Fig.\ref{fig:uprim}, that  amounts to set 
\begin{equation}   
u'(\delta \theta)=   \frac{1}{2} u'(0).
     \label{eq:upo}
\end{equation} 
Putting this expression in the solution  (\ref{eq:solG}) equivalent to  $u'(\theta)= u'(0)(1+ \theta^{2}/  \theta_{c}^{2})^{-1/\alpha}$,  with $\theta_{c}=\delta \theta \, \tilde \theta_{c}$, we get
\begin{equation}   
\tilde \theta_{c}^{2}=   \frac{1}{2^{\alpha}-1} 
     \label{eq:tetac}
\end{equation} 
that yields a quantitative expression for the relation  (\ref{eq:scalingangle})
\begin{equation}   
\delta \theta ^{2}=  \eta \,\left( \tilde \gamma {\mathcal I}(0)\, \frac{4}{\tilde \theta_{c}^{2} }\frac{1-\alpha}{\alpha} \frac{c_{\alpha}} {d_{\alpha}}\right)
     \label{eq:finalscale-app}
\end{equation} 
 where 
all coefficients in the parenthesis are numerical ones and dimensionless, $ \theta_{c} $ , $c_{\alpha}$ and $d_{\alpha}$ are defined  just above, $ {\mathcal I} (0) $ is deduced from  (\ref{eq:defI2}), 
 \begin{equation} 
 {\mathcal I} (0) = \frac{2}{1-\alpha} \int _0^{\infty}  {\mathrm{d}} \zeta   \frac {1}  {\zeta^{\alpha} (1-\zeta)}, 
    \label{eq:defI0}
\end{equation}   
and the coefficient $\tilde \gamma$  in front of the integral defining $\tilde \sigma_{ij}^{Re} $ is arbitrary, but must have an opposite sign with respect to $ {\mathcal I} (0) $, namely must be negative for $\alpha < 1/2$ and positive for $\alpha > 1/2$, as illustrated  in Fig.\ref{fig:I0-coeff}-(a) plotting $ {\mathcal I} (0) $ versus $\alpha$.    Fig.\ref{fig:I0-coeff}-(b) displays the ratio $\delta \theta ^{2} / (\eta \tilde \gamma {\mathcal I}(0)) $  versus $\alpha$, see (\ref{eq:finalscale}).

The velocity component  $u$ can now be expressed from (\ref{eq:up2}) and the component $v $ from 
 \begin{equation} 
 v'(\theta) = -\frac{\eta U}{\delta \theta}  \tilde g''_1(\tilde \theta) \theta
    \label{eq:vp2}
\end{equation}   
where we set $\sin \theta \approx \theta$. It follows that $v$ is of order   $\eta^{3/2}$, although $u$ is bigger, of order $\eta$ (the integration over $\theta$ amounts to multiply the prefactor $\eta U/\delta \theta$ by $\delta \theta$).
Inserting in (\ref{eq:solGapp})  the relations (\ref{eq:bound}), (\ref{eq:tetac}), and defining $k= \delta \theta ^{2} / \eta$ in (\ref{eq:finalscale}), the integration of (\ref{eq:up2}) and (\ref{eq:vp2}) gives the velocity profiles
 \begin{equation} 
u(\theta) -U = \frac{\eta U}{d_{\alpha}} \int _0^{\theta/\theta_{c}}  {\mathrm{d}}y \, (1+y^{2})^{-1/\alpha}
    \label{eq:profu-app}
\end{equation}   
and
 \begin{equation} 
v(\theta) -v(0)=  \frac{ \eta U \theta_{c}} {d_{\alpha}} \int _0^{\theta/\theta_{c}}  {\mathrm{d}}y \, y (1+y^{2})^{-1/\alpha}
    \label{eq:profv-app1}
\end{equation}   
where $\theta_{c}=  \tilde \theta_{c} \delta \theta =  \tilde \theta_{c} \sqrt{k \eta}$, which gives 
 \begin{equation} 
v(\theta) = \,-\,  \frac{ \eta^{3/2}U k^{1/2} \tilde \theta_{c}}{d_{\alpha}}\,\frac{\alpha}{2(1-\alpha)}\left (1+\frac{\theta^{2}}{\theta_{c}^{2}}\right )^{-\frac{1-\alpha} {\alpha}},
    \label{eq:profv-app2}
\end{equation}   
where $$ k=\tilde \gamma {\mathcal I}(0)\, \frac{4}{\tilde \theta_{c}^{2} }\frac{1-\alpha}{\alpha} \frac{c_{\alpha}} {d_{\alpha}},$$  is a positive constant  because the product $\tilde \gamma {\mathcal I}(0)$ has to be positive.

\section{Comparison with the mixing layer experiment  of  K. Sodjovi.}
\label{sec:Rennes}
Let us  first recall that  when using the  Boussinesq model  for turbulent stress tensor, ($\sigma^{Re} =\nu_{t}u_{,y}$  where the turbulent viscosity $\nu_{t}$ is a linear fonction of $x$ , independent of $y$), one found that  the width of the turbulent domain scales as $\eta$ for small values of this ratio $\eta$ \cite{Schlichting}-\cite{these-Rennes}. In the literature  we found few experimental works  devoted to  small values of $\eta$,  most of them being concerned  by ratios  $U_{1}/U_{2} $ of order few units. 
 Nevertheless we  found  in  \cite{these-Rennes} a measurement  extending from $\eta=0.05$ up to $\eta=0.6$
 which  displays a peculiar non linear behavior of the curve $\delta \theta$ versus $\eta$, at small $\eta$ values,
  although the author concludes that the width of the turbulent domain grows linearly with $\eta$ in agreement with the Boussinesq model. 
   
   \begin{figure}
\centerline{ 
(a)  \includegraphics[height=1.5in]{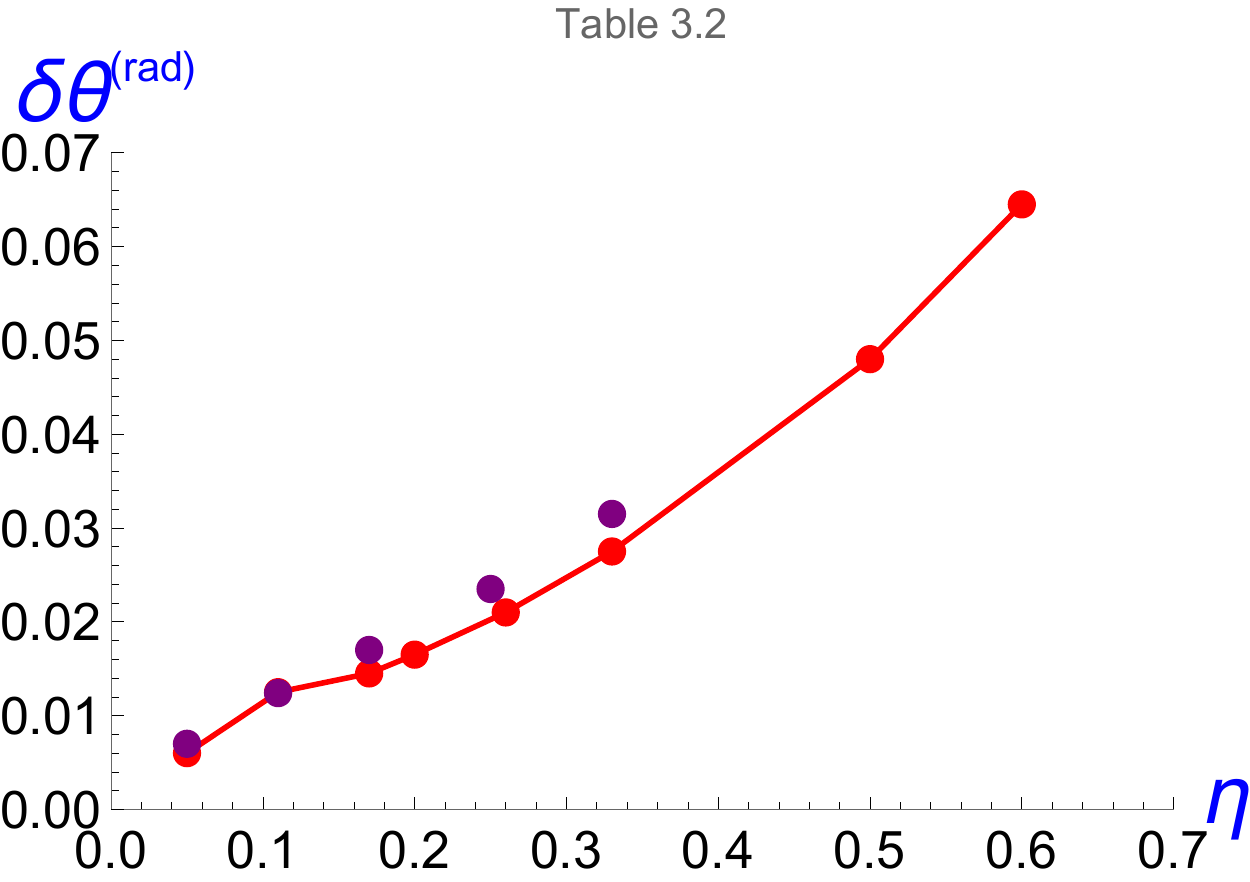}
(b) \includegraphics[height=1.5in]{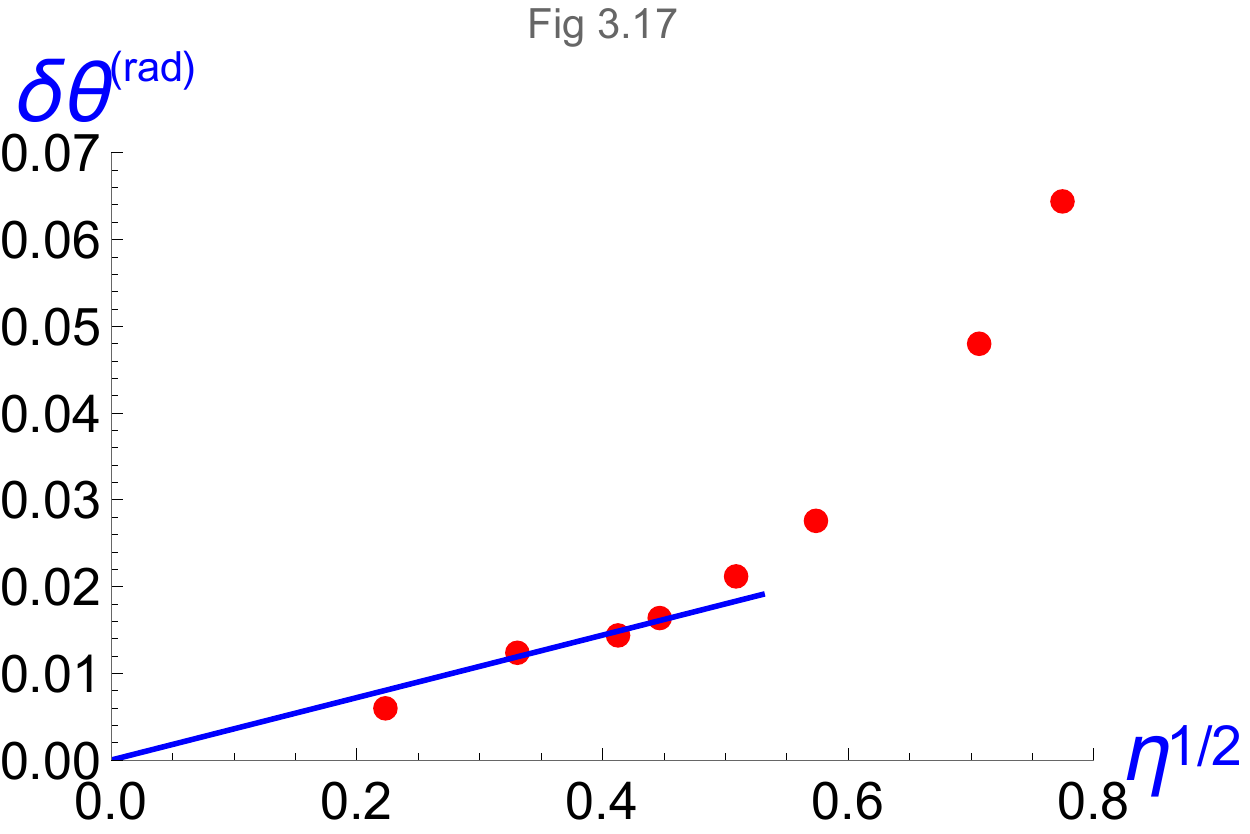}
  }
\caption{  Angular width $\delta \theta$ from  experimental data  summarized in Table $3.2$ of \cite{these-Rennes} :  (a) Half angle of the turbulent domain as a function of  $\eta$, (b)  Same measure as a function of $\eta^{1/2}$.  In (a) The  red points on the solid line correspond to measurements made by the author of the  thesis \cite{these-Rennes} which surprisingly agree with the data of  \cite{mehta}  (purple points) made over a shorter range,  the latter being enhanced by a factor 2.  Note that the two  data display the same behavior.   In (b) the data  reasonably agree with the prediction of our model, $\delta \theta \sim \eta^{1/2}$ for small values of $\eta$ and even medium values. The blue line  of (b) is drawn  simply to compare with our prediction.
}
\label{fig:Rennes}
\end{figure}

Using the data published in table 3.2 of the thesis \cite{these-Rennes}
 we have plotted $\delta \theta$ versus $\eta ^{1/2}$, see Fig.\ref{fig:Rennes}.  It appears that  the plot agrees quite well with  our prediction (\ref{eq:scaling}),  for $ \eta $ smaller than $0.26$ , which is  not a small domain ($\eta=0.26$ corresponds to $U_{1}= 1.7 U_{2}$).
 
\begin{figure}
\centerline{ 
  \includegraphics[height=1.5in]{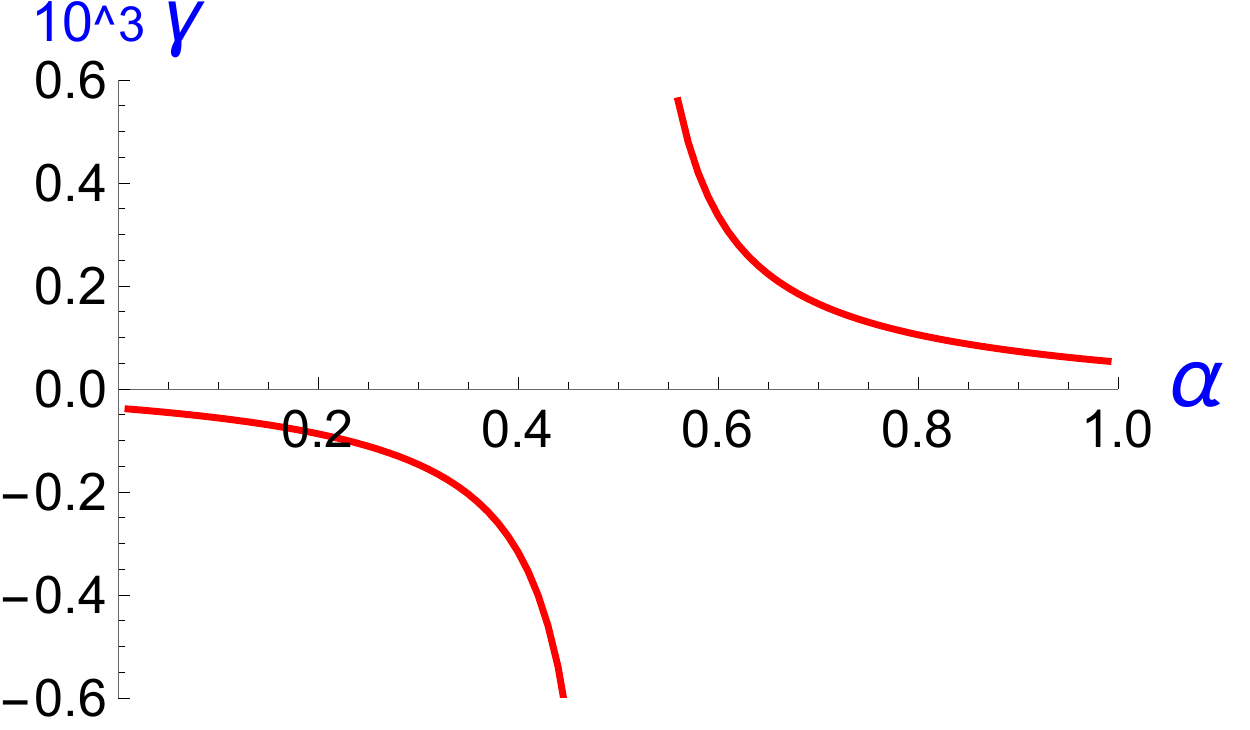}
  }
\caption{  $10^{3}\tilde \gamma$  versus the exponent $\alpha$. The numerical value of the prefactor $\tilde \gamma$ is deduced from the data of \cite{these-Rennes}  and from the relation (\ref{eq:tildegam}). Recall that  in a small  domain close to $\alpha=1/2$, the approximation $ {\mathcal I}(\theta - \theta') \approx  {\mathcal I}(0) $ is not valid because ${\mathcal I}(0) =0$, see Fig. \ref{fig:Ixi}-(a).
}
\label{fig:tildegam}
\end{figure}

 From figure \ref{fig:Rennes}  we deduce the  experimental value of the ratio $\delta \theta/\sqrt{\eta} $. The experiment provides $\sqrt{k}= 0.036$ defined in (\ref{eq:finalscale}) or  (\ref{eq:finalscale-app}),  that allows to obtain  a numerical value for the   prefactor $\tilde \gamma$ in (\ref{eq:sig11})
   \begin{equation} 
 \tilde \gamma(\alpha)  = k \frac{1-\alpha}{\alpha} \tilde \theta_{c}^{2} \frac{d_{\alpha}}{c_{\alpha}} \frac{1}{4 {\mathcal I}(0)}.
   \label{eq:tildegam}
\end{equation} 
It  depends on the exponent $\alpha$, as  illustrated  in Fig.\ref{fig:tildegam}, see captions for the divergence for  $\alpha=1/2$.  Close to this peculiar value, one may use the approximation ${\mathcal I}(\theta - \theta') \approx {\mathcal I}(\delta \theta)$, or solve  more precisely the equation $C+D=0$, something not done here.

 \section{ order of magnitude of   $\tilde \sigma_{ij}^{Re} $    }
\label{sec:app-sigI}
From equation (\ref{eq:sigfin}) and the hypothesis $\mathcal I (\theta-\theta')= {\mathcal I}(0)$,  we have 
\begin{equation} 
\tilde \sigma_{xx}^{Re} \, =\,  \tilde \gamma \rho {\mathcal I}(0) \vert g+g'' (\theta) \vert  ^{1- \alpha}  \int {\mathrm{d}} \theta'   \,  \vert g+g'' (\theta) \vert  ^{ \alpha} \,  \tilde  \tau_{xx} (\theta') 
\label{eq:appDI1}
\end{equation} 
where $ \tilde  \tau_{xx}(\theta') = (g+g'') \sin 2\theta' $, see (\ref{eq:tauxx}). Because $  \tau_{xx}= \,-\, \tau_{yy} $ we have
\begin{equation} 
\tilde \sigma_{xx}^{Re} \, =\,  - \, \tilde \sigma_{yy}^{Re}.
\label{eq:appDI2}
\end{equation} 
The non diagonal element of the tensor  $\tilde \sigma_{ij}^{Re} $  is 
\begin{equation} 
\tilde \sigma_{xy}^{Re} \, =\,  \tilde \gamma \rho {\mathcal I}(0) \vert g+g'' (\theta) \vert  ^{1- \alpha}  \int {\mathrm{d}} \theta'   \,  \vert g+g'' (\theta) \vert  ^{ \alpha} \,  \tilde  \tau_{xy}(\theta') 
\label{eq:appDI3}
\end{equation} 
where  $ \tilde  \tau_{xy}(\theta') = -(g+g'') \cos 2\theta' $ is an even function. Using the same argument as above, we have $ \tilde \sigma_{xy}^{Re} \sim \delta \theta (g'')^{2} $ or $\tilde \sigma_{xy}^{Re}  \, \sim \, \eta^{3/2} U^{2}$. The integral in (\ref{eq:appDI3}) does not vanish because the integrand is an even function, then
\begin{equation} 
\tilde \sigma_{xy}^{Re}  \sim \eta^{3/2},
\label{eq:appDI4}
\end{equation} 
A priori the diagonal elements  $ \tilde \sigma_{xx}^{Re} = - \tilde  \sigma_{yy}^{Re} $   are  of order $(\delta \theta)^{2}(g'')^{2}  \sim \eta^{2} U^{2} $ under the condition that the integrand is even. But $\tilde \tau_{xx} $ is an odd function, then we have $\tilde \sigma_{xx}^{Re} \,= - \tilde  \sigma_{yy}^{Re} =0 $ at order $ \eta^{2}$. To go further we have to 
notice that the results  of this section are obtained within the rough approximation $ {\mathcal I}(\theta-\theta')  =  {\mathcal I}(0) $ which greatly simplifies the calculation.  This approximation is valid  for small values of the angles, except for $\alpha = 1/2$ where  ${\mathcal I}(0) =0$. This point is considered below.

\subsection{ Case $\alpha \neq 1/2$ }
More generally,  close to $\theta=\theta'$, for any $\alpha$ values  we have shown in (\ref{eq:Ixi}) that 
\begin{equation} 
{\mathcal I}(\theta-\theta')  \approx   {\mathcal I}(0) + b \vert \theta-\theta' \vert,
\label{eq:Ixib}
\end{equation} 
where $b $ is a factor  quasi-independent of $\alpha$ ($b\approx 9$), whereas $  {\mathcal I}(0)  $ changes a lot with $\alpha$ (more precisely it grows from $-5$ up to $25$ when $\alpha$  increases from $-1/4 $ to $3/4$, see Fig.\ref{fig:Ixi}-b).
The second term in (\ref{eq:Ixib}) changes the above results as follows: all odd  terms which have been considered as   providing a null contribution to the integral over $\theta'$ in $\sigma_{ij}^{Re}$, will now bring a non zero contribution and provide a component  having an order of magnitude equal to the same order of magnitude as before times $ b \delta \theta \sim \eta^{1/2}$ (the approximate value of  $ {\mathcal I}(\theta-\theta') $) for small argument.   In summary, for  $\alpha \neq 1/2$  we  get
\begin{equation}
 \left \{ \begin{array}{l}
\tilde  \sigma_{xx}^{Re}  =\, - \, \tilde \sigma_{yy}^{Re} \,\sim \eta^{5/2}\\
\tilde   \sigma_{xy}^{Re}  \, \sim  \eta^{3/2}
  \mathrm{,}
\end{array}
\right \}
\label{eq:corsigI}
\end{equation}
which confirms that the tensor $\tilde  \sigma_{ij}^{Re}$ does not satisfy the realizability conditions
\begin{equation}
 \left \{ \begin{array}{l}
 \sigma_{ii}  \geqslant 0 \\
  \sigma_{ij} ^{2} \, \leqslant \sigma_{ii} \sigma_{jj}  \qquad {\text {for}}  \qquad i  \neq j
  \mathrm{,}
\end{array}
\right \}
\label{eq:appDcondreal}
\end{equation}
Indeed  both conditions in  (\ref{eq:appDcondreal}) are  not fulfilled,  one diagonal element  of $\tilde  \sigma$ is negative, moreover (\ref{eq:corsigI})  shows that the order of magnitude of the components are  inconsistent with the Schwarz inequality, a  binding constraint of the Reynolds stress defined by (\ref{eq:sigRe1}).

\subsection{ Case $\alpha=1/2$ }
\label{sec:appE}
In the case $\alpha=1/2$, one has $ {\mathcal I}(0)=0  $ and  $ {\mathcal I}(\theta-\theta') \approx\, b \sqrt{1-\cos(\theta-\theta')}  \sim b \vert \theta - \theta' \vert $ where $b \approx 8.9 $.
For the tensor $\sigma_{xx}^{I}$, the components are respectively of order
\begin{equation}
 \left \{ \begin{array}{l}
\tilde \sigma_{xx}^{Re} = -  \tilde \sigma_{yy}^{Re} \,\sim \, {\mathcal I}(\delta \theta) (g'') ^{2} \delta \theta \,\sin \theta \,  \sim \,   \eta^{5/2}  U^{2}   \\
\tilde   \sigma_{xy}^{Re}  \, \sim  {\mathcal I}(\delta \theta) (g'')^{2}   \delta \theta\, \sim \,   \eta^{2}  U^{2}  
  \mathrm{.}
\end{array}
\right. \label{eq:sigI1s2}
\end{equation}
which  does not satisfy the realizability conditions  because the second equation yields  $ \sigma_{xy} ^{2} \; > \;  \sigma_{xx} \,\sigma_{yy}  $ as in the general case of $0 < \alpha <1$.  But the realizability conditions are satisfied  by adding the diagonal tensor (\ref{eq:sig22})    to the tensor $\tilde \sigma_{ij}^{Re}$ because $$ \sigma_{ii}^{Re,p} \sim (\eta U)^2, $$   which is  positive, of the same order as $\tilde \sigma_{xy}^{Re}$, and satisfy the Schwarz inequality when taking $\gamma= \tilde \gamma$ as in the general case.

  \end{document}